\documentclass[a4paper,11pt]{iopart}
\usepackage{bm,amssymb,epsfig,cite,color}
\newcommand{\be}{\begin{equation}}
\newcommand{\ee}{\end{equation}}
\newcommand{\bea}{\begin{eqnarray}}
\newcommand{\eea}{\end{eqnarray}}

\def\bra#1{|#1\rangle}

\def\g{\gamma}
\def\b{\beta}
\def\d{\delta}
\def\a{\alpha}
\def\e{\varepsilon}

\def\nn{\nonumber\\}

\def\r#1{(\ref{#1})}

\def\2t#1#2{\langle\tau_{#1}\tau_{#2}\rangle}

\def\nn{\nonumber\\}
\def\tt{\tilde{\theta}}
\def\fr#1{(\ref{#1})}

\def\i{{\rm i}}
\def\d{{\rm d}}
\def\e{{\rm e}}
\def\dps{\displaystyle}

\eqnobysec
\begin{document}

\title[Slowest relaxation mode of the PASEP with open
boundaries]{Slowest relaxation mode of the partially asymmetric
exclusion process with open boundaries}
\author{Jan de Gier$^1$ and Fabian H L Essler $^2$}
\address{$^1$ Department of Mathematics and Statistics, The University of
  Melbourne, 3010 VIC, Australia\\
$^2$ Rudolf Peierls Centre for Theoretical Physics, University
  of Oxford, 1 Keble Road, Oxford, OX1 3NP, United Kingdom}

\begin{abstract}
We analyze the Bethe ansatz equations describing the complete spectrum
of the transition matrix of the partially asymmetric exclusion process
on a finite lattice and with the most general open boundary conditions. We extend results 
obtained recently for totally asymmetric diffusion [J. de Gier and
F.H.L. Essler, J. Stat. Mech. P12011 (2006)] to the case of partial
asymmetry. We determine the finite-size scaling of the spectral gap,
which characterizes the  approach to stationarity at late times, in
the low and high density regimes and on the coexistence line. We
observe boundary induced crossovers and discuss possible
interpretations of our results in terms of effective domain wall
theories.   
\end{abstract}


\maketitle

\section{Introduction}

The partially asymmetric simple exclusion process (PASEP) \cite{ASEP1,ASEP2}
is one of the most thoroughly studied models of non-equilibrium
statistical mechanics \cite{Derrida98,Schuetz00,GolinelliMallick06,BEreview}. 
It is a microscopic
model of a driven system \cite{schmittmann} describing the asymmetric
diffusion of hard-core particles along a one-dimensional chain with
$L$ sites. At late times the PASEP exhibits a relaxation towards a
non-equilibrium stationary state. In the presence of two boundaries at
which particles are injected and extracted with given rates, the bulk
behaviour at stationarity is strongly dependent on the injection and
extraction rates. The corresponding phase diagram as well as various
physical quantities have been determined by exact methods
\cite{Derrida98,Schuetz00,DEHP,gunter,sandow,EsslerR95,PASEPstat1,PASEPstat2,BEreview}.

Given the behaviour in the stationary state an obvious question is how
the system relaxes to this state at late times. For the PASEP on a ring, where
particle number is conserved, such results were obtained by means
of Bethe's ansatz some time ago \cite{dhar,BAring1,BAring2}.
More recently there has been considerable progress in analyzing the
dynamics in the limit of totally asymmetric exclusion and on an infinite lattice, see e.g.
\cite{praehoferS02a, praehoferS02b, ferrariS06,sasamoto07a ,sasamoto07b, borodin}, where random matrix techniques can be used.  

For the finite system with open boundaries there have been several studies of dynamical
properties by means of numerical, phenomenological and renormalization
group methods \cite{numerics1,numerics2,numerics3,KSKS,DudzS00,HaSt06}. 
For the case of symmetric diffusion a Bethe ansatz solution was
constructed in \cite{robin}.
Recently, we have applied Bethe's ansatz to the PASEP with open
boundaries \cite{GierE05,GierE06}. It is well-known that the PASEP
can be mapped onto the spin-1/2 anisotropic Heisenberg chain with
general (integrable) open boundary conditions
\cite{sandow,EsslerR95}. By building on recent progress in applying
Bethe's ansatz to the latter problem 
\cite{Cao03,Nepo02,NepoR03,GierP04,Nepo05,YangNZ} we determined the Bethe
ansatz equations for the PASEP with the most general open boundary
conditions. By analyzing these equations we derived the finite size
scaling behaviour of the spectrum of low-lying excited states for the
cases of symmetric and totally asymmetric diffusion. Upon varying the
boundary rates, we observed crossovers in massive regions, with
dynamic exponents $z=0$, and between massive and scaling regions with
diffusive ($z=2$) and KPZ ($z=3/2$) behaviour. 

In the present work we extend these results to the case of partially
asymmetric diffusion, where the analysis of the spectrum is
significantly more involved.

\begin{figure}[ht]
\centerline{
\begin{picture}(320,80)
\put(0,0){\epsfig{width=0.7\textwidth,file=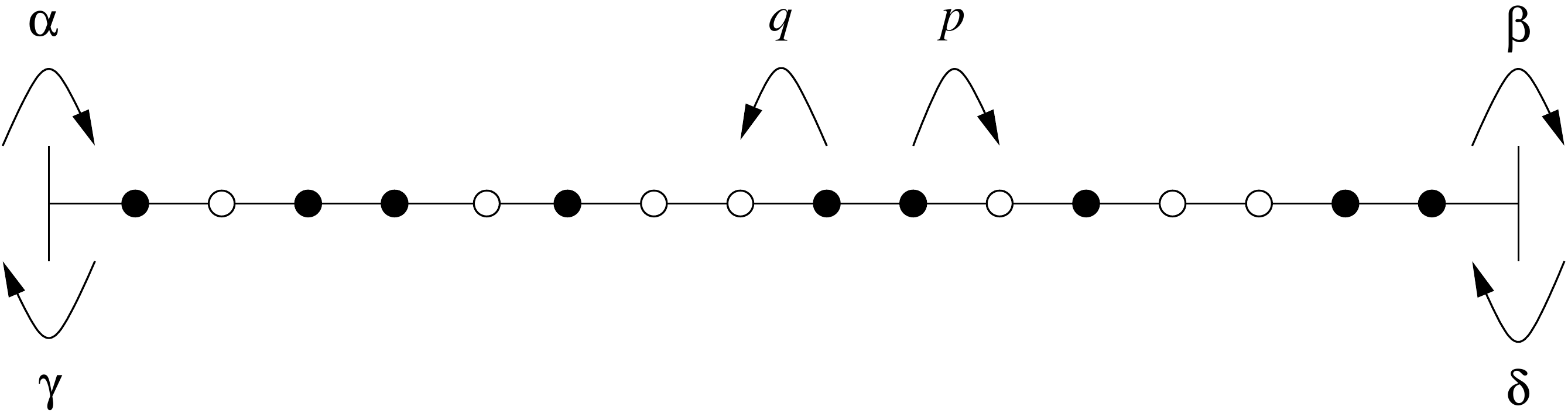}}
\end{picture}}
\caption{Transition rates for the partially asymmetric exclusion process.}
\label{fig:paseprules}
\end{figure}

We now turn to a description of the dynamical rules defining the
PASEP on a one dimensional lattice with $L$ sites. At any given time
$t$ each site is either occupied by a
particle or empty. The system is then updated as follows. In the bulk
of the system, i.e. sites $i=2,\ldots,L-1$, a particle attempts to hop
one site to the right with rate $p$ and one site to the left with rate
$q$. The hop is prohibited if the neighbouring site is occupied. On
the first and last sites these rules are modified. If site $i=1$ is
empty, a particle may enter the system with rate $\alpha$. If on the
other hand site $1$ is occupied by a particle, the latter will leave
the system with rate $\gamma$. Similarly, at $i=L$ particles are
injected and extracted with rates $\delta$ and
$\beta$ respectively. 

It is customary to associate a Boolean variable $\tau_i$ with every
site, indicating whether a particle is present ($\tau_i=1$) or not
($\tau_i=0$) at site $i$. Let $\bra0$ and $\bra1$ denote the standard
basis vectors in $\mathbb{C}^2$. A state of the system 
at time $t$ is then characterized by the probability distribution
\be
\bra{P(t)} = \sum_{\bm \tau} P(\bm{\tau}|t) \bra{\bm{\tau}},
\ee
where
\be
\bra{\bm{\tau}} = \bra{\tau_1,\ldots,\tau_L} = \bigotimes_{i=1}^{L} \bra{\tau_i}.
\ee
The time evolution of $\bra{P(t)}$ is governed by the aforementioned
rules, which gives rise to the master equation
\bea
\frac{{\rm d}}{{\rm d} t} \bra{P(t)} &=& M \bra{P(t)},
\label{eq:Markov}
\eea
where the PASEP transition matrix $M$ consists of two-body
interactions only and is given by 
\be
M = \sum_{k=1}^{L-1} I^{(k-1)}\otimes\widetilde{M}\otimes I^{(L-k-1)}
+ m_1 \otimes I^{(L-1)}+  I^{(L-1)}\otimes m_L.
\label{eq:TransitionMatrix}
\ee
Here $I^{(k)}$ is the identity matrix on the k-fold tensor product
of $\mathbb{C}^2$ and $\widetilde{M}: \mathbb{C}^2\otimes \mathbb{C}^2
\rightarrow  \mathbb{C}^2\otimes \mathbb{C}^2$ is given by
\be
\widetilde{M} = \left(\matrix{
0 & 0 & 0 & 0\cr
0 & -q & p & 0 \cr
0 & q & -p & 0 \cr
0 & 0 & 0 & 0}\right).
\ee
The terms involving $m_1$ and $m_L$ describe injection
(extraction) of particles with rates $\a$ and $\delta$ ($\g$ and $\b$) at
sites $1$ and $L$ respectively. Their explicit forms are
\be
m_1=\left(\matrix{-\a&\g \cr \a &-\g\cr}\right),\qquad 
m_L=\left(\matrix{-\delta&\b\cr \delta&-\b\cr}\right).
\label{h1l}
\ee

The transition matrix $M$ has a unique stationary state
corresponding to the eigenvalue zero. For positive rates, all
other eigenvalues of $M$ have non-positive real parts. The late time
behaviour of the PASEP is dominated by the eigenstates of $M$ with the
largest real parts of the corresponding eigenvalues, as follows from the following argument.  
The average of an observable $X$ is given by (see e.g. Ref. \cite{Vlad})
\be
\langle X\rangle(t)=\langle 0| X \e^{Mt}|P_0\rangle
\ee
Here $P_0$ is an initial state and $\langle 0|$ is the left eigenstate of $M$ with eigenvalue 0. This may be written in the
spectral representation with respect to the eigenstates of $M$
\be
\langle X\rangle(t)=\sum_n\langle 0| X|n\rangle  \e^{\mathcal{E}_nt} a_n,
\ee
where $|P_0\rangle=\sum_n a_n |n\rangle$ and
$M|n\rangle=\mathcal{E}_n|n\rangle$. In the limit $t\to\infty$ only
the stationary state survives and we have 
\be
\lim_{t\to\infty} \langle X\rangle(t)=\langle 0| X|0\rangle  a_0.
\ee
At very late times we have
\be
\langle X\rangle(t)\approx\langle 0| X|0\rangle  a_0+\langle
0|X|1\rangle a_1 e^{\mathcal{E}_1t}, 
\ee
where ${\mathcal E}_1$ is the eigenvalue of $M$ with the largest real
part that contributes to the spectral decomposition of the initial
state $|P_0\rangle$. Hence the relaxation rate $-\mathcal{E}_1$
determines the approach to the stationary state at asymptotically late
times. In the next
sections we determine the eigenvalue of $M$ with the largest non-zero
real part using Bethe's ansatz. The latter reduces the problem of
determining the spectrum of $M$ to solving a system of coupled
polynomial equations of degree $3L-1$. Using these equations, the
spectrum of $M$ can be studied numerically for very large $L$, and, as
we will show, analytic results can be obtained in the limit
$L\rightarrow\infty$. 

\section{Bethe ansatz equations}
In \cite{GierE05,GierE06} it was shown that the PASEP transition
matrix $M$ can be diagonalised using the Bethe Ansatz. In
\cite{GierE06} the Bethe equations were analyzed in some detail
for the case $q=\gamma=\delta=0$ corresponding to totally asymmetric
diffusion (TASEP). The resulting TASEP dynamical phase diagram
displays interesting crossovers within the low and high density
phases, at which the transition matrix eigenvalue corresponding to the
slowest relaxation mode changes non-analytically.

This work is concerned with the generalization of some of these
results to the PASEP case with $q,\gamma,\delta\neq 0$. Before
turning to the technical details of our analysis we present a summary
of our main results. Throughout this work we set without loss of
generality $q<p=1$. For simplicity we only consider $L$ even, as for 
odd $L$ the details will be somewhat different. 
 
As was shown in \cite{GierE05,GierE06}, all eigenvalues ${\cal E}$ of
$M$ can be expressed in terms of the roots $z_j$ of a set of
$L-1$ non-linear algebraic equations as \footnote{We rescale the roots
  $z_j$ in eqns (3.1) - (3.3) of \cite{GierE06} by $Q=\sqrt{q}$.} 
\bea
{\cal E}= -\mathcal{E}_0-\sum_{j=1}^{L-1}\frac{\left(q-1\right)^2
  z_j}{(1-z_j)(qz_j-1)},
\label{eq:pasep_en}
\eea
where
\be
\mathcal{E}_0 = \alpha+\beta+\gamma+\delta.
\ee
The complex roots $z_j$ satisfy the Bethe ansatz equations
\bea
\left[\frac{qz_j-1}{1-z_j}\right]^{2L} K(z_j) =\prod_{l\neq j}^{L-1}
\frac{qz_j-z_l}{z_j-qz_l} \frac{q^2z_jz_l-1} {z_jz_l-1},\
j=1\ldots L-1.\nn
\label{eq:pasep_eq}
\eea
Here $K(z) = \tilde{K}(z,\alpha,\gamma) \tilde{K}(z,\beta,\delta)$, where
\be
\tilde{K}(z,\alpha,\gamma) =
\frac{(z+\kappa^+_{\alpha,\gamma})
  (z+\kappa^-_{\alpha,\gamma})}{(q\kappa^+_{\alpha,\gamma}z+1)
  (q\kappa^-_{\alpha,\gamma}z+1)},
\ee
and
\bea
\kappa^{\pm}_{\alpha,\gamma} &=& \frac{1}{2\alpha} \left(
v_{\alpha,\gamma} \pm \sqrt{v_{\alpha,\gamma}^2 +4\alpha\gamma}\right),\\
v_{\alpha,\gamma} &=& 1-q-\alpha+\gamma.
\eea
In order to ease notations we will use the following abbreviations, 
\be
a=\kappa^+_{\alpha,\gamma},\quad b=\kappa^+_{\beta,\delta},\quad c=\kappa^-_{\alpha,\gamma},\quad d=\kappa^-_{\beta,\delta}.
\label{eq:ab}
\ee
The constant $\mathcal{E}_0$ is expressed in our new notations as
\be
\mathcal{E}_0 = (1-q)\left(\frac{1-ac}{(1+a)(1+c)}
+\frac{1-bd}{(1+b)(1+d)}\right).
\label{eq:e0}
\ee

\section{Lowest excitation of the PASEP in the ``forward-bias
  regime'': summary of main results}
By analysing the set of equations (\ref{eq:pasep_eq}) for large,
finite $L$ we have determined the eigenvalue of the transition matrix
with the largest non-zero real part. From this ``lowest excited state
energy'' we can infer properties of the relaxation towards the stationary
state at asymptotically late times. In the present work we have
restricted our analysis to the regime of small values of the
parameters $c$ and $d$. This corresponds loosely to a ``forward-bias
regime'' in which particles diffuse predominantly from left to right
and particle injection and extraction occurs mainly at sites $1$ and
$L$ respectively. The restrictions on the permitted values of $c$ and $d$
are discussed in more detail in section \ref{se:cd}.

\subsection{Stationary state phase diagram}
\begin{figure}[ht]
\centerline{\includegraphics[width=200pt]{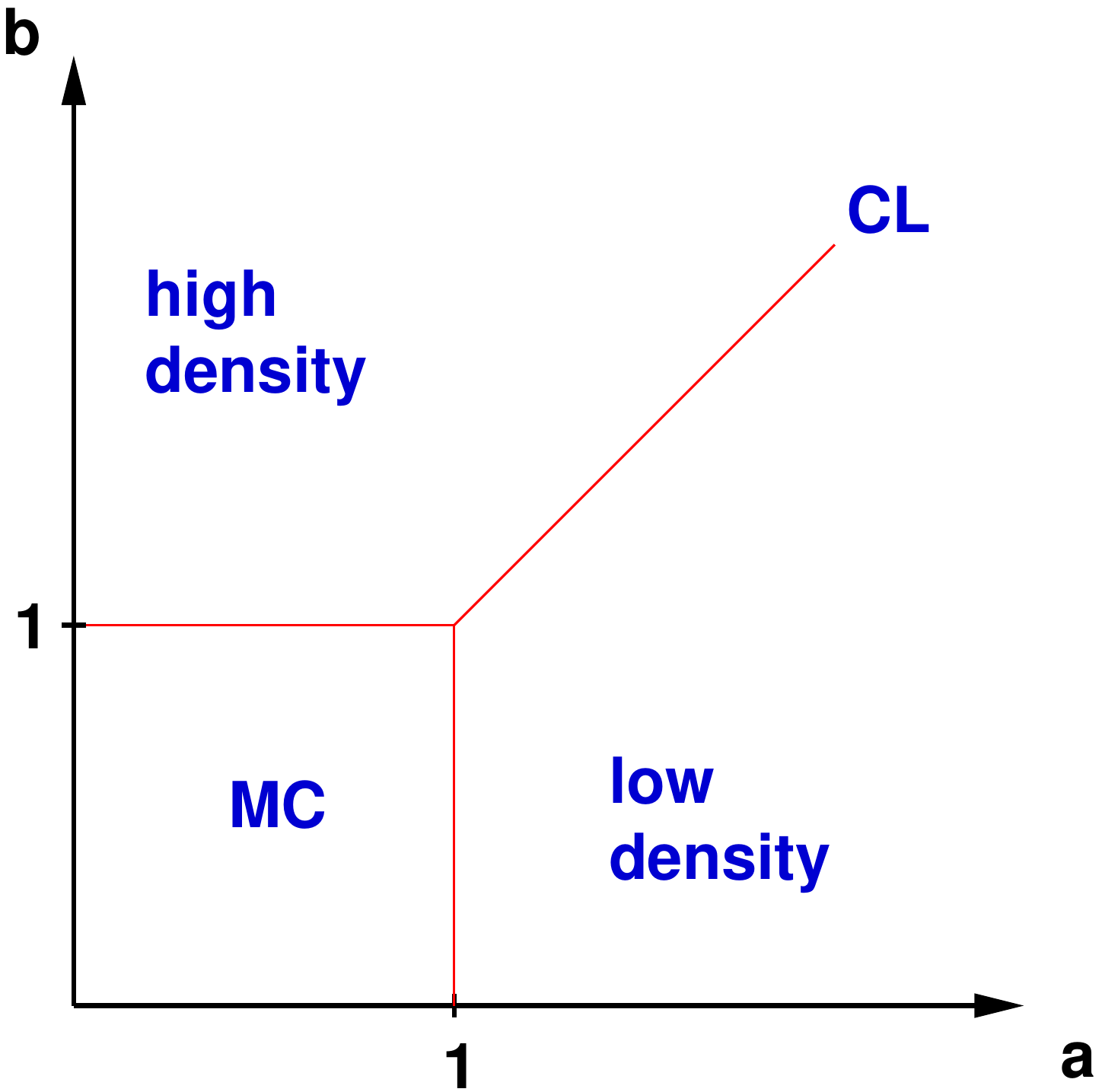}}
\caption{Stationary state phase diagram of the PASEP.
The high and low density phases are separated by the
coexistence line (CL). The maximum current phase (MC) occurs at small
values of the parameters $a$ and $b$ defined in (\ref{eq:ab}). } 
\label{fig:statPD}
\end{figure}

The phase diagram of the PASEP at stationarity was found by Sandow
\cite{sandow} and is depicted in Figure~\ref{fig:statPD}. We note that
the phases depend only on the parameters $a$ and $b$ defined in
(\ref{eq:ab}) rather than $p,q,\alpha,\beta,\gamma,\delta$ separately.

\subsection{Dynamical phase diagram}

\begin{figure}[ht]
\centerline{\includegraphics[width=250pt]{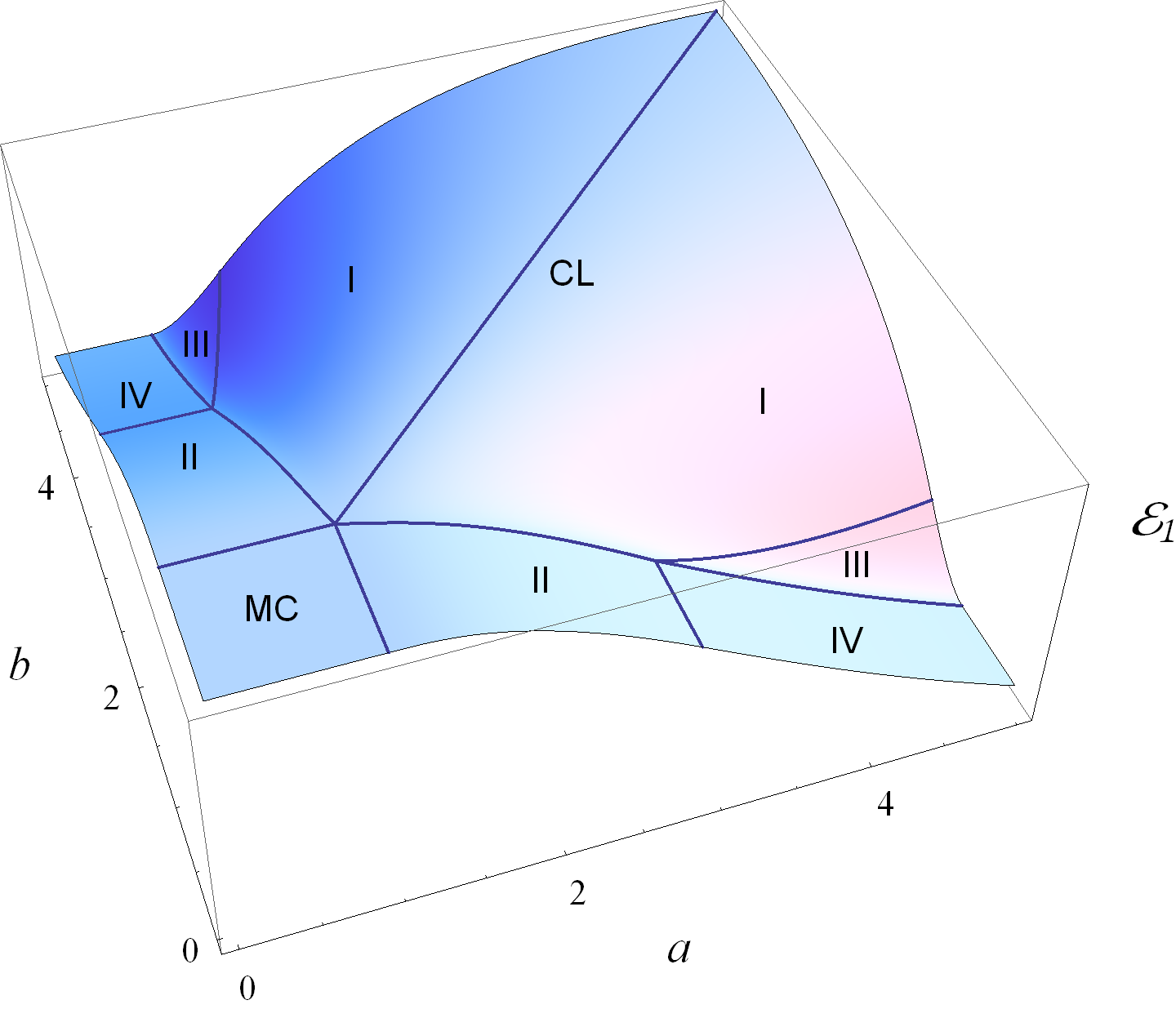}}
\caption{Dynamical phase diagram of the PASEP in the forward bias
regime determined by the lowest excitation $\mathcal{E}_1$. The
horizontal axes are the boundary parameters $a$ and $b$ \fr{eq:ab} and
the vertical axis is the lowest relaxation rate ${\cal E}_1$. 
The latter goes to
zero for large systems on the coexistence line (CL) and in the maximum
current phase (MC). The curves and lines correspond to various
crossovers in the low and high density phase, across which
$\mathcal{E}_1$ changes   non-analytically. See the main text for a
detailed explanation.}  
\label{fig:phasediagram}
\end{figure}

The dynamical phase diagram for the PASEP resulting from our analysis in the
regime $q<p=1$ is shown in Figure~\ref{fig:phasediagram}. It exhibits
the same subdivision into low and high density phases ($a>1$ and $b>1$), 
the coexistence line ($a=b>1$) and the maximum current phase
($a,b<1$) found from the analysis of the current in the stationary
state \cite{sandow}. However, the finite-size scaling of the
lowest excited state energy of the transition matrix suggests the sub-division
of both low and high-density phases into four regions
respectively. These regions are characterized by different functional
forms of the relaxation rates at asymptotically late times.
We will describe the four regions in the low density phase ($a>1$ and
$b\leq a$). The results in the high density phase are obtained by exchanging
$a\leftrightarrow b$. We note that the results for the relaxation
rates presented below are valid in the limit $L\to\infty$ at fixed $q<1$.
In particular the limit of symmetric exclusion $q=1$ cannot be
obtained by taking the limit $q\to 1$ in the expression presented
below.

\begin{enumerate}
\item {\bf Region I}:

This region is defined by
\be
b \geq q^2a\  {\rm for}\  a>q^{-3/2}\ {\rm and}\  b \geq a^{-1/3}\ {\rm for}\ 1<a<q^{-3/2}.
\label{eq:curves}
\ee
The eigenvalue of the lowest excitation is given by ($q<1$ fixed)
\bea
\mathcal{E}_1 &=& -(1-q)\left[\frac{1}{1+a} + \frac{1}{1+b} 
+ \frac{2z_{\rm c}}{1-z_{\rm c}}\right]\nonumber\\
&&+\frac{(1-q)}{L^2} \frac{\pi^2}{(z_{\rm c}^{-1}-z_{\rm c})}+{\cal O}(L^{-3}),
\label{E_I}
\eea
where
\be
z_{\rm c}=-\frac{1}{\sqrt{ab}}.
\ee

\item {\bf Coexistence Line:} 

The coexistence line is defined by $a=b>1$ and separates the low and high
density phases. We find that the leading term in \fr{E_I} vanishes and that
the lowest eigenvalue concomitantly scales with the system size as
\be
\mathcal{E}_1 = \frac{1-q}{L^2} \frac{\pi^2}{(a^{-1}-a)} + \mathcal{O}(L^{-3}),
\label{E_CL}
\ee
The inverse proportionality of the eigenvalue \fr{E_CL} 
to the square of the system size implies a dynamic exponent
$z=2$, which in turn suggests that the relaxation at late times is
governed by diffusive behaviour. 
\item {\bf Region II:} 

This region is defined by
\be
b \leq a^{-1/3}\ {\rm for}\ 1<a<q^{-3/2}\ .
\ee
The eigenvalue of the lowest excitation is now independent of $b$
\bea
\mathcal{E}_1 &=& -(1-q)\left[\frac{1}{1+a} + \frac{2z_{\rm
      c}+1}{1-z_{\rm c}}\right]\nonumber\\
&&+ \frac{1-q}{L^2} \frac{4\pi^2}{(z_{\rm c}^{-1}-z_{\rm c})}
+{\cal O}(L^{-3}),
\label{E_II}
\eea
where
\be
z_{\rm c}=-a^{-1/3}.
\ee

We note that the leading terms of $\fr{E_I}$ and $\fr{E_II}$ coincide
along the boundary $b=a^{-1/3}$ separating the two regimes, but the 
terms of order $L^{-2}$ exhibit a discontinuity. This suggests a
crossing of levels and an associated change in the detailed nature
of the corresponding relaxational mode.
\item {\bf Region III:} 

This region is defined by
\be
q^{1/2}\leq b \leq q^2a\ {\rm for}\  a>q^{-3/2}.
\ee
Up to terms of order $\mathcal{O}(L^{-3})$, the eigenvalue of the
lowest excitation in this region is given by 
\be
\mathcal{E}_1 \approx -(1-q)\left(\frac{1}{1+a} + \frac{1}{1+b} + \frac{q a}{1+q a} + \frac{2z_{\rm c}-1}{1-z_{\rm c}} \right),
\label{E_III}
\ee
where now
\be
z_{\rm c}=-q/b.
\ee
We note that the leading terms of $\fr{E_I}$ and $\fr{E_III}$ coincide
along the boundary $b=q^2a$ separating Regions I and III. However,
throughout Region III there is no contribution of order ${\cal
  O}(L^{-2})$ to the transition matrix eigenvalue of the lowest
excited state.
\item {\bf Region IV:}

This final region is defined by
\be
a>q^{-3/2},\qquad b<q^{1/2}.
\ee
The eigenvalue of the lowest excitation in this region is given by
\bea
\mathcal{E}_1&=& -(1-q)\left[
\frac{1}{1+a} + \frac{q a}{1+q a} + \frac{2z_{\rm c}}{1-z_{\rm
    c}}\right]\nonumber\\
&&+\frac{1-q}{L^2} \frac{\pi^2}{(z_{\rm c}^{-1}-z_{\rm c})}+{\cal O}(L^{-3}),
\label{E_IV}
\eea
where now
\be
z_{\rm c}=-q^{1/2}.
\ee
We note that the leading terms of \fr{E_IV} and \fr{E_III} match along
the boundary between regions IV and III. The same holds for the
leading terms of \fr{E_IV} and \fr{E_II} along the boundary between
regions IV and II. On the other hand, there is a discontinuity in the
${\cal O}(L^{-2})$ contributions in both cases.
\end{enumerate}

\subsection{Modified domain wall theory}
\label{se:domain}

It was shown in \cite{KSKS,DudzS00} that
the diffusive relaxation towards the stationary state found on the
coexistence line as well as in the low and high density phases can be
understood in terms of an effective domain wall theory (DWT). In this
approach the excited states driving the relaxational dynamics are
modelled as domain walls between low and high density regions. They
carry out a random walk with right and left hopping rates given by
\be
D_\pm = (1-q)\frac{\rho^\pm(1-\rho^\pm)}{\rho^+ -\rho^-}.
\label{Drates}
\ee
Here, $\rho^- = 1/(1+a)$ and $\rho^+ = b/(1+b)$ are the stationary
bulk densities in the low and high-density phases respectively. The
domain walls are assumed to be reflected from both
boundaries. Interestingly, domain wall theory gives the exact 
stationary state along the curve $ab=q^{-1}$ in parameter space
\cite{KFS}. It is furthermore possible to construct an entire family
of exact domain wall solutions of the master equation \cite{KFS}
\footnote{It was shown in Ref. \cite{KFS} that
exact multi domain wall solutions of the master equation exist
more generally along the curves $ab=q^{-n}$ for $n=1,2,3,\ldots$, but
their precise properties have been analyzed only for $n=1$ 
\cite{gunterPC}.}.
The leading relaxation rate calculated from these exact domain wall
solutions agrees with our results \fr{E_I} and \fr{E_CL} in Region I
and the coexistence line. This suggests that DWT gives a correct
description of the relaxational behaviour at late times throughout
these regimes.

In contrast, the eigenvalue of the transition matrix
determined from DWT does not coincide with our results in Regions
II-IV \footnote{For the case of totally asymmetric diffusion this was
already observed numerically in Ref.~\cite{numerics1} and analytically
in Ref.~\cite{GierE06}.}. This means
that while the shock profile considered in \cite{KFS} remains the
exact stationary state along the curve $abq=1$, 
the slowest relaxational mode is no longer given by the particular
implementation of DWT proposed in \cite{KSKS,DudzS00}. An
obvious question is whether it is possible to reproduce our findings
by a suitably modified DWT. 

\begin{figure}[ht]
\centerline{\includegraphics[width=200pt]{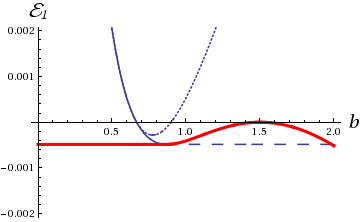}}
\caption{The lowest excitation for $q=0.6$ and $a=1.5$. The functions
  defined in \fr{E_I}, \fr{E_III} (dotted) and  \fr{E_II} (dashed) are
  both displayed for $0<b<2$. The excitation gap is a combination of
  \fr{E_I} and \fr{E_II}, and drawn as the bold curve (red online).
} 
\label{fig:cross1}
\end{figure}

To this end it is useful to consider our results for the eigenvalue of
the lowest excited state as a function of $b$ for fixed $a$ and $q$.
A particular example ($a=1.5$ and $q=0.6$) is shown in
Fig.~\ref{fig:cross1}.

When $b>a$ we are in the high density phase and the system is gapped.
The corresponding gap is finite and given by \fr{E_I}. Decreasing $b$
we are approaching the coexistence line $b=a$, where the gap
vanishes and the relaxation is purely diffusive. Decreasing $b$
further drives us into the low density phase and the gap is again
finite. We expect the gap to grow as $b$ decreases. However, at
$b_{\rm c}=a^{-1/3}$ ($\approx 0.87$ in the example shown in 
Fig.~\ref{fig:cross1}), the slope of \fr{E_I} as a function of
$b$ vanishes and for $b<b_{\rm c}$ \fr{E_I} increases with decreasing $b$.
For values of $b$ smaller than the crossover point $b_{\rm c}$, the gap is
no longer described by \fr{E_I} but by \fr{E_II}, and remains constant.

It is now straightforward to reproduce these results within the
framework of an effective DWT. In Region I the DWT prediction
\cite{KFS} for the gap coincides with \fr{E_I} 
\be
\mathcal{E}_1(\rho^-, \rho^+) = -D_+ - D_- + 2\sqrt{D_+D_-}\ ,
\ee
where $D_\pm$ are defined above. However, as we cross over into Region
II the gap predicted by this DWT no longer agrees with the exact result.
We therefore modify the DWT as follows. We postulate that in Region II
the density $\rho^+$ ceases to depend on $b$ and remains fixed at
$\rho^+_{\rm eff} = b_{\rm c}/1+b_{\rm c}$. Retaining the expressions \fr{Drates} for
the hopping rates of the domain wall one finds that the gap is then
given by $\mathcal{E}_1(\rho^-,\rho^+_{\rm   eff})$. The value of
$\rho^+_{\rm   eff}$ is determined from the requirement that 
\be
\left.\frac{\partial \mathcal{E}_1(\rho^-, \rho^+)}{\partial \rho^+}
\right|_{\rho^+=\rho^+_{\rm eff}} = 0.  
\ee
By construction this modified DWT reproduces the exact result for the
relaxation rate.

\begin{figure}[ht]
\centerline{\includegraphics[width=200pt]{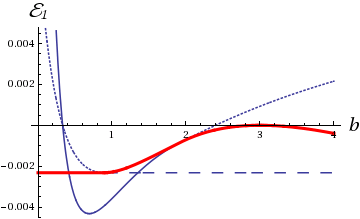}} 
\caption{Eigenvalue of the lowest excitation for $q=0.8$ and
  $a=3$. The functions defined in \fr{E_I}, \fr{E_III} (dotted) and
  \fr{E_IV} (dashed) are displayed for $0<b<4$. The excitation gap is
  a combination of all three, and drawn as the bold curve (red
  online).
} 
\label{fig:cross2}
\end{figure}

The modification of the DWT becomes more involved if
$a>q^{-3/2}$. In this case there are two crossovers as is shown in
Fig.~\ref{fig:cross2} for the particular example $q=0.8$ and $a=3$.
The first crossover separating Regions I and III takes place at
$b_{{\rm c},1}=q^{2}a$ ($\approx 1.92$ for the example shown in
Fig.~\ref{fig:cross2}). For values of $b<b_{{\rm c},1}$ the gap is given by 
\fr{E_III}. In this case the $\mathcal{O}(L^{-2})$ correction to
$\mathcal{E}_1$ vanishes. A second crossover, now between Regions III
and IV, occurs at $b_{{\rm c},2}=q^{1/2}$ ($\approx0.89$ in the example shown
in Fig.~\ref{fig:cross2}) where   
\be
\left.\frac{\partial \mathcal{E}_1(\rho^-, \rho^+)}{\partial \rho^+}
\right|_{\rho^+=\rho^+_{\rm eff}} = 0.
\ee
For values of $b<b_{{\rm c},2}$ the relaxation rate is given by \fr{E_III}
and no longer depends on $b$. 

It is clearly possible to reproduce the exact relaxation rates by adjusting the 
densities $\rho^\pm$ in the DWT accordingly. Unlike above, in this case we do not have a convincing heuristic argument for the first crossover. The modifications of DWT described here are completely {\sl ad hoc}, all we can say is that at the boundaries between the various regions, levels cross and the precise nature of the relaxational dynamics changes. It would be very interesting to
investigate whether in Regions II-IV the relevant excited states are still domain walls.

\section{Analysis of the Bethe ansatz equations}
\label{se:BAanalysis}
\begin{figure}[ht]
\centerline{
\begin{picture}(320,100)
\put(0,0){\epsfig{width=150pt,file=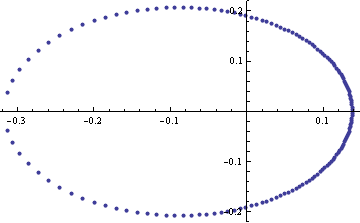}}
\put(170,0){\epsfig{width=150pt,file=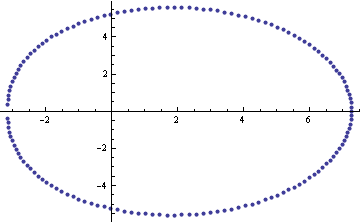}}
\end{picture}}
\caption{Root distribution and reciprocal root distributions for $a=\kappa^+_{\alpha,\gamma}=5$, $b=\kappa^+_{\beta,\delta}=2$, $c=\kappa^-_{\alpha,\gamma}=-0.01$, $d=\kappa^-_{\beta,\delta}=-0.023$, $q=0.1$ and $L=150$} 
\label{fig:roots}
\end{figure}

In the following we derive the results summarised in the previous
section. To this end we analyse (\ref{eq:pasep_en}) and
(\ref{eq:pasep_eq}) in the limit or large lattice lengths $L$.
It is convenient to introduce functions
\bea
g_{\rm }(z) &=& \ln \left( \frac{z(1-qz)^2}{(z-1)^2}\right),\label{eq:g}\\
g_{\rm b}(z) &=& \ln \left(\frac{z(1-q^2z^2)}{1-z^2}\right) +
\ln\left(\frac {z+a}{1+qaz}\frac{1+c/z}{1+qcz}\right) \nonumber\\
&& {}+
\ln\left(\frac {z+b}{1+qbz}\frac{1+d/z}{1+qdz}\right).\label{eq:gb}
\eea
The central object of our analysis is the ``counting function''
\cite{yaya69,deVegaW85,book},  
\be
\i Y_L(z) = g_{\rm}(z) + \frac{1}{L} g_{\rm b}(z)
+ \frac{1}{L} \sum_{l=1}^{L-1} K(z_l,z),
\label{eq:logtasepBAE}
\ee
where $K(w,z)$ is given by
\be
K(w,z) = -\ln \left( \frac{w-q z}{1-qw/z} \frac{1-q^2zw}{1-w z}\right).
\label{eq:KernelDef}
\ee
Using the counting function, the Bethe ansatz equations
\fr{eq:pasep_eq} can be cast in logarithmic form as
\be
Y_L(z_j) = \frac{2\pi}{L} I_j\ ,\qquad j=1,\ldots,L-1.
\label{eq:Z=I}
\ee
Here $I_j$ are integer numbers. Each set of integers $\{I_j|\; j=1,\ldots,
L-1\}$ in (\ref{eq:Z=I}) specifies a particular (excited) eigenstate
of the transition matrix. Based on numerical solutions of (\ref{eq:Z=I}) 
using standard root finding techniques, we assume that the first excited 
state always corresponds to the same set of integers
\be
I_j = -L/2+j\quad {\rm for}\quad j=1,\ldots,L-1.
\label{eq:Idef}
\ee
The corresponding roots lie on a simple curve in the complex plane,
which approaches a closed contour as $L\rightarrow \infty$. The
latter fact is more easily appreciated by considering the locus of
reciprocal roots $z_j^{-1}$ rather than the locus of roots $z_j$.
In Fig.~\ref{fig:roots} we present results for $a=\kappa^+_{\alpha,\gamma}=5$, $b=\kappa^+_{\beta,\delta}=2$, $c=\kappa^-_{\alpha,\gamma}=-0.01$, $d=\kappa^-_{\beta,\delta}=-0.023$, $q=0.1$ and $L=150$.

\begin{figure}[ht]
\begin{center}
\begin{picture}(180,115)
\put(0,-20){\epsfig{width=180pt,file=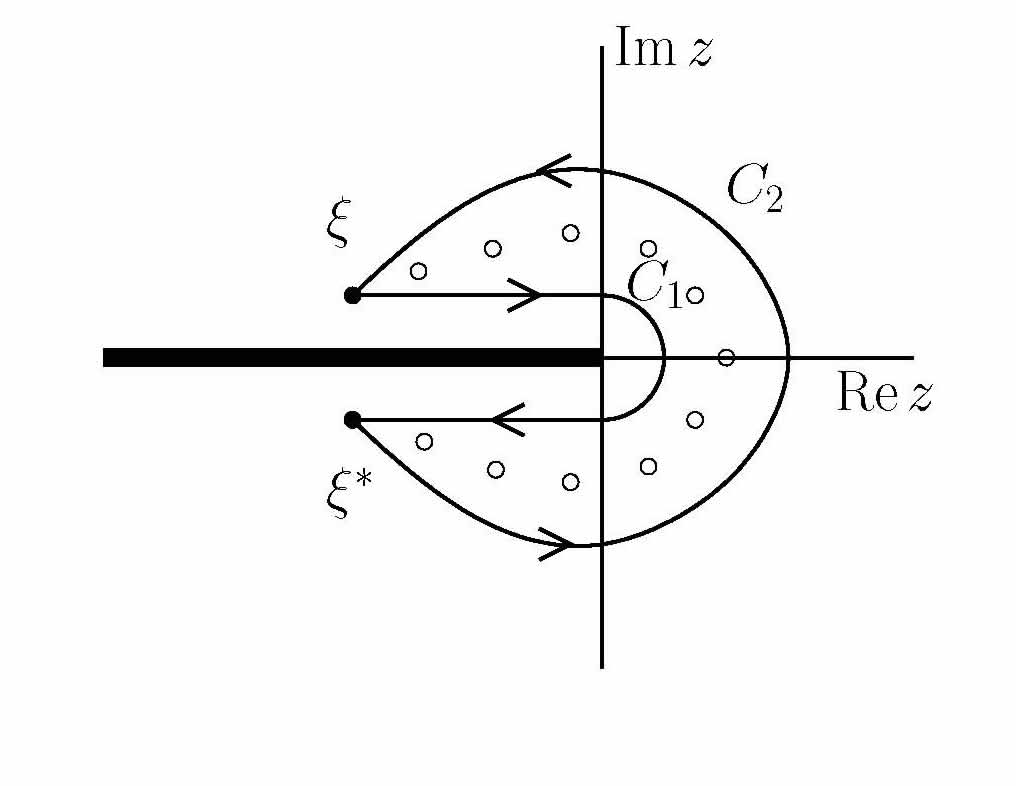}}
\end{picture}
\end{center}
\caption{Sketch of the contour of integration $C$ in \fr{eq:sum2int}. The
  open dots correspond to the roots $z_j$ and $\xi$ is chosen close to
  $z_{L-1}$ and avoiding poles of $\cot(LY_L(z)/2)$.} 
\label{fig:contour}
\end{figure}

In order to compute the exact large $L$ asymptotics of the spectral
gap, we derive an integro-differential equation for the counting
function $Y_L(z)$ in the limit $L\rightarrow\infty$. As a simple
consequence of the residue theorem we can write 
\be
\frac1L \sum_{j=1}^{L-1} f(z_j) = \oint_{C_1+C_2} 
\frac{dz}{4\pi\i}\ f(z)
Y_L'(z) \cot\left(\frac12 L Y_L(z)\right),
\label{eq:sum2int}
\ee
where $C=C_1+C_2$ is a contour enclosing all the roots $z_j$, $C_1$
being the ``interior'' and $C_2$ the ``exterior'' part, see
Fig.~\ref{fig:contour}. The contours $C_1$ and $C_2$ intersect in
appropriately chosen points $\xi$ and $\xi^*$. It is convenient to fix
the end points $\xi$ and $\xi^*$ by the requirement
\be
Y_L(\xi^*) = -\pi +\frac{\pi}{L},\qquad Y_L(\xi) = \pi -\frac{\pi}{L}.
\label{eq:xidef}
\ee
Using (\ref{eq:xidef}) in (\ref{eq:logtasepBAE}) we obtain a
nonlinear integro-differential equation for the counting function
$Y_L(z)$. Our goal is to solve this equation for large lattice lengths
$L$ through an expansion in inverse powers of $L$. In order to do so
we first rewrite \r{eq:sum2int} by separating the contributions coming
from $C_1$ and $C_2$. When doing this it is useful to note that on the
contour of integration we have by definition of the counting function
that ${\rm Im} Y_L(z)=0$. As a result the imaginary part of $Y_L(z)$ is
positive on $C_1$ and negative on $C_2$. Using the fact that
integration from $\xi^*$ to $\xi$ over the contour formed by the roots
is equal to half that over $C_2 - C_1$ we find, 
\bea
\i\,Y_L(z) &=& g(z) + \frac{1}{L} g_{\rm b}(z) +\frac{1}{2\pi}
\int_{\xi^*}^{\xi} K(w,z) Y'_L(w) \d w \nonumber \\ 
&& \hspace{-1cm}+ \frac{1}{2\pi} \int_{C_1} \frac{K(w,z)Y'_L(w)}{1-\e^{-\i L
Y_L(w)}}\, \d w + \frac{1}{2\pi} \int_{C_2} \frac{K(w,z)Y'_L(w)}{
\e^{\i L Y_L(w)}-1}\,\d w,
\label{eq:intY}
\eea
where we have chosen the branch cut of $K(w,z)$ to lie along the
negative real axis. 

Our strategy is to solve the integro-differential equation \fr{eq:intY}
by iteration. Once we have constructed the desired solution $Y_L(z)$
we determine the corresponding eigenvalue of the transition matrix
from equation \fr{eq:pasep_en} by turning the sum over roots into an
integral using \fr{eq:sum2int}
\bea
\mathcal{E} &=& -\mathcal{E}_0 -\frac{L}{2\pi} \int_{\xi^*}^{\xi} \varepsilon(z) Y_L'(z)\, \d z \nonumber\\
&& - \frac{L}{2\pi} \int_{C_1} \frac{\varepsilon(z)Y'_L(z)}{1-\e^{-\i L
Y_L(z)}}\, \d z - \frac{L}{2\pi} \int_{C_2} \frac{\varepsilon(z)Y'_L(z)}{
\e^{\i L Y_L(z)}-1}\,\d z.
\label{eq:intE}
\eea
Here the constant ${\mathcal E}_0$ was defined previously in
\fr{eq:e0} and the ``bare energy'' $\varepsilon(z)$ is
\bea
\varepsilon(z) &=& \frac{(q-1)^2 z}{(1-z)(qz-1)}
=(1-q)\left(\frac{1}{z-1}-\frac{1}{qz-1}\right). 
\label{eq:epsdef}
\eea

\section{Low and High Density Phases}
\label{se:LandH}

The low density phase for $q<1$ is characterized by $a>1$ and $b<1$,
while the high density phase corresponds to $a<1$ and $b>1$. In these
phases we find that the locations of the end points $\xi$ and $\xi^*$
are such that a straightforward expansion of the correction term in
\fr{eq:intY} in inverse powers of $L$ is possible (see
e.g. \cite{Olver,PovoPH03} and \ref{app:abel}). The result is 
\bea
\i\,Y_L(z) &=& g(z) + \frac{1}{L} g_{\rm b}(z) +\frac{1}{2\pi}
\int_{\xi^*}^{\xi} K(w,z) Y'_L(w) \d w \nonumber\\
&& {} + \frac{\pi}{12L^2} \left(
\frac{K'(\xi^*,z)}{Y'_L(\xi^*)} - \frac{K'(\xi,z)}{Y'_L(\xi)}\right) +
     \mathcal{O}(L^{-4}),
\eea
where the derivatives of $K$ are with respect to the first
argument. We note that here we have implicitly assumed that
$Y'_L(\xi)$ is nonzero and of order ${\cal O}(L^0)$. The integral from
$\xi^*$ to $\xi$ is along the contour formed by the roots, see
Fig.~\ref{fig:contour}. In order to utilize complex analysis techniques
it is useful to extend the integration contour beyond the endpoints,
so that it pinches the negative real axis at points $z_{\rm c}^\pm=z_{\rm c}\pm
\i 0$ ($z_{\rm c}\in {\mathbb R}$). This leads to the following expression
\bea
\i\,Y_L(z) &=& g(z) + \frac{1}{L} g_{\rm b}(z) +\frac{1}{2\pi}
\int_{z_{\rm c}^-}^{z_{\rm c}^+} K(w,z) Y'_L(w) \d w \nonumber\\
&&{} + \frac1{2\pi} \int_{\xi^*}^{z_{\rm c}^-} K(w,z) Y'_L(w) \d w + \frac1{2\pi} \int_{z_{\rm c}^+}^{\xi} K(w,z) Y'_L(w) \d w \nonumber\\
&& {} + \frac{\pi}{12L^2} \left(
\frac{K'(\xi^*,z)}{Y'_L(\xi^*)} - \frac{K'(\xi,z)}{Y'_L(\xi)}\right) +
     \mathcal{O}(L^{-4})\ .
\label{eq:intY_M}
\eea
The key to the solution of \fr{eq:intY_M} is that all terms have
simple expansions in inverse powers of $L$. In order to find the
eigenvalue \fr{eq:intE} up to order ${\cal O}(L^{-2})$ we need to
solve \fr{eq:intY_M} to order ${\cal O}(L^{-3})$. Substituting the
expansions  
\be
Y_L(z) = \sum_{n=0}^\infty L^{-n} y_n(z),\qquad \xi = z_{\rm c} +
\sum_{n=1}^\infty L^{-n} (\delta_n + \i \eta_n),
\label{eq:expansion}
\ee
back into (\ref{eq:intY_M}) yields a hierarchy of integro-differential
equations for the functions $y_n(z)$
\be
y_n(z) = g_n(z) + \frac{1}{2\pi \i} \int_{z_{\rm c}^-}^{z_{\rm c}^+}
K(w,z) y'_n(w) \,\d w.
\label{eq:yn}
\ee
The integral is along the closed contour following the locus of the
roots, see Fig.~\ref{fig:contour}. The first few driving terms $g_n(z)$
are given by 
\be
\renewcommand{\arraystretch}{1.2}
\begin{array}{r@{\hspace{2pt}}c@{\hspace{3pt}}l}
g_0(z) &=& -\i g(z), \\
g_1(z) &=& -\i g_{\rm b}(z) + \kappa_1 + \lambda_1 \tilde{K}(z_{\rm c},z),
\\
g_2(z) &=& \kappa_2 + \lambda_2 \tilde{K}(z_{\rm c},z) + \mu_2 K'(z_{\rm
  c},z), \\
g_3(z) &=& \kappa_3 + \lambda_3 \tilde{K}(z_{\rm c},z) + \mu_3 K'(z_{\rm
  c},z) + \nu_3 K''(z_{\rm c},z).
\end{array}
\label{eq:gs}
\ee
The functions $g$ and $g_{\rm b}$ are defined in \fr{eq:g} and
\fr{eq:gb} and  
\be
\tilde{K}(z_{\rm c},z) = -\ln(-z_{\rm c})+ \ln \left( \frac{1-qz_{\rm c}z^{-1}}{1-q zz_{\rm c}^{-1}} \frac{1-zz_{\rm c}}{1-q^2zz_{\rm c}}\right).
\ee
The terms involving the kernel and its derivatives arise from 
Taylor-expanding the integrands in the integrals from $\xi^*$ to
$z_{\rm c}^-$  and from $z_{\rm c}^+$ to $\xi$. Concomitantly the coefficients
$\kappa_n$, $\lambda_n$, $\mu_n$ and $\nu_n$ are given in terms of
$\delta_n$, $\eta_n$ defined by \fr{eq:expansion}, and by
derivatives of $y_n$ evaluated at $z_{\rm c}$. Explicit expressions
are presented in \ref{ap:Mcoef}. We show how to construct a general
solution of the set of equations \fr{eq:yn} for $n\leq 3$ under
certain  restrictions on the values of the parameter $a$, $b$, $c$ and
$d$ in \ref{ap:countingfunction}. 
Having this solution in hand, we may determine the coefficients
$\kappa_n$, $\lambda_n$, $\mu_n$ and $\nu_n$ as follows. Substituting
the expansions \fr{eq:expansion} into the boundary condition
\fr{eq:xidef}, which fixes the endpoints $\xi$ and $\xi^*$, we obtain
a hierarchy of conditions for $y_n(z_{\rm c})$, e.g. 
\bea
Y_L(\xi)&=&y_0(\xi)+\frac{1}{L}y_1(\xi)+\frac{1}{L^2}y_2(\xi)+\ldots\nn
&=&y_0(z_{\rm c})+\frac{1}{L}\left[y_1(z_{\rm c})+y_0'(z_{\rm c})(\delta_1+i\eta_1)\right]+\ldots\nn
&=&\pi-\frac{\pi}{L}\ .
\label{eq:yexpand}
\eea
Solving this equation order by order, we find that in all cases
considered in the present work
\be
\renewcommand{\arraystretch}{1.2}
\begin{array}{r@{\hspace{2pt}}c@{\hspace{3pt}}l}
\lambda_3 &=& \mu_2 = \lambda_2 = \kappa_1=0,\\
\nu_3 &=& \dps z_{\rm c} \mu_3 = \frac{\lambda_1}{2\i} z_{\rm c} (z_{\rm c}-1)^2 \kappa_2 =
\frac{\pi^2 \lambda_1 (1+\lambda_1^2)} {6} \frac{z_{\rm c}^2(1-z_{\rm c})^2}{(1+z_{\rm c})^2}.
\end{array}
\label{eq:constants}
\ee
Here the parameter $\lambda_1$ depends on the values of $a$ and
$b$. As will be described in detail below, it determines various
crossover regimes within the low and high density phase.  

Having determined the counting function we may use equation
\fr{eq:intE} to evaluate the corresponding eigenvalue of the
transition matrix. Evaluating the necessary integrals in the same way
as for the counting function itself we obtain
\be
{\cal E} = -\mathcal{E}_0 - \frac{L}{2\pi} \oint_{z_{\rm c}} \varepsilon(z)Y'_L(z)\d z -\i \sum_{n\geq 0} e_n L^{-n},
\label{eq:energyMI}
\ee
where the integral is over the closed contour on which the roots lie,
$\varepsilon$ is given in \fr{eq:epsdef} and 
\bea
e_0 &=& \lambda_1 \varepsilon(z_{\rm c}),\nonumber\\
e_1 &=& \lambda_2 \varepsilon(z_{\rm c}) + \mu_2 \varepsilon'(z_{\rm c}),\\
e_2 &=& \lambda_3 \varepsilon(z_{\rm c}) + \mu_3 \varepsilon'(z_{\rm c}) + \nu_3 \varepsilon''(z_{\rm c}).
\nonumber
\eea
Substituting
the expansion for $Y_L(z)$ in inverse powers of $L$ into
\fr{eq:energyMI} we arrive at the following result for the eigenvalue
of the transition matrix with the largest non-zero real part
\bea
\mathcal{E}_1 &=& -(1-q)\left(\frac{1}{1+a} + \frac{1}{1+b} +
\frac{2z_{\rm c}-2-\i\lambda_1}{1-z_{\rm c}} -\sum_m \frac{c_m}{1-z_m}
\right) \nonumber\\ 
&& {}+ \frac{1}{L^2} \frac{(1-q)\,\i\lambda_1
  (1+\lambda_1^2)\pi^2}{6(z_{\rm c}^{-1}-z_{\rm c})}
+\mathcal{O}(L^{-3}). 
\label{eq:M_E}
\eea
Here, the sum over
$m$ is over all poles of $g'_{\rm b}(z)$ other than $0$, $-c$ and
$-d$, that lie inside the contour of integration. The constants $c_m$
are the corresponding residues. We note that the number and position
of such poles depend on the values of the parameters $a$ and 
$b$. The values of both $\lambda_1$ and $z_{\rm c}$ in turn depend on
these poles. In particular we find 
\be
\lambda_1 = 2\i + \i\sum_m c_m\ .
\ee
The result \fr{eq:M_E} for the smallest relaxation rate is generically
a constant of order ${\cal O}(L^0)$, implying an exponentially fast
relaxation to the stationary state at large times. We note that due to
the symmetry of the root distribution corresponding to \fr{eq:Idef}
under complex conjugation, ${\cal E}_1$ is in fact real, and hence
there are no oscillations in the slowest relaxation mode. 

\subsection{Region I: large values of $a$ and $b$}
%
The first regime we consider is obtained loosely speaking by taking
$q$ to be small, $a$ and $b$ large and positive, $c$ and $d$ small and
negative. More precisely we require
\begin{enumerate}
\item $-a$ and $-b$ lie outside the contour of integration in
  \fr{eq:intY_M}, 
\item $-1/qa$ and $-1/qb$ lie outside the contour,
\item $-c$ and $-d$ lie inside the contour of integration.
\end{enumerate}
We will assume that the last assumption is fulfilled, postponing a
detailed discussion to Section~\ref{se:cd}. Condition (i) amounts to
the inequalities  $-a < z_{\rm c}$ and $-b < z_{\rm c}$, which
translate to
\be
b> b_{\rm c,1}=a^{-1/3} \quad (a>1),\qquad\quad
a> a_{\rm c,1}= b^{-1/3} \quad (b>1).
\label{eq:curves1}
\ee
Condition (ii) implies that $-1/qa < z_{\rm c}$ and $-1/qb < z_{\rm
  c}$, resulting in
\be
b< b_{\rm c,2}=q^2 a \quad (a>1),\qquad\qquad a< a_{\rm c,2}=q^2 b
  \quad (b>1). 
\label{eq:lines1}
\ee
From the
distribution of the reciprocal roots, see Fig.~\ref{fig:roots}, we infer
that for these values of the parameters, the roots lie in fact inside
the unit circle. We therefore assume, and verify a posteriori, that
$z_{\rm c}\neq -1$ and that the points $\pm 1$ lie outside the contour
of integration. 
Combining the above assumption we conclude that the driving term 
\fr{eq:gs} for \fr{eq:yn} with $n=1$ can be represented in the form
\bea
g_1(z) &=& -\i\ln z -\i \ln \left(\frac{z+c}{z}\right)-\i \ln
\left(\frac{z+d}{z}\right)\nonumber\\ 
&& {} +\lambda_1 \ln \left(\frac{z-qz_{\rm c}}{z}\right) +g_1^{\rm a}(z),
\eea
where $g_1^{\rm a}(z)$ is analytic inside the contour of
integration. Under the above assumptions we may now solve the system
\fr{eq:yn} of integro-differential equations and then verify a
posteriori that all underlying assumptions in fact hold.
Some details of this caculation are presented in
\ref{ap:countingfunction}. The result for the counting function is
$Y_L(z)=y_0(z)+\frac{1}{L}y_1(z)+\frac{1}{L^2}y_2(z)+\frac{1}{L^3}y_3(z)+{\cal
  O}(L^{-4})$ where
\bea
y_{0}(z) &=& -\i \ln \left[ - \frac{z}{z_{\rm c}}
\left(\frac{1-z_{\rm c}}{1-z}\right)^2 \right],
\label{eq:Z0sol1}\\
y_1(z) &=& -\i\ln\left[ -\frac{z}{z_{\rm c}} \frac{1-z_{\rm c}^2}{1-z^2}\right] +\kappa_1 - \i\ln\left(ab\right) -\lambda_1 \ln(-z_{\rm c}) \nonumber\\
&&{} -\i\ln\left[\frac{(-c/z;q)_\infty (-cz;q)_\infty (-z/a;q)_\infty (-qaz_{\rm c};q)_\infty} {(-c/z_{\rm c};q)_\infty (-cz_{\rm c};q)_\infty (-z_{\rm c}/a;q)_\infty (-qaz;q)_\infty}\right] \nonumber\\
&&{}-\i\ln\left[\frac{(-d/z;q)_\infty (-dz;q)_\infty (-z/b;q)_\infty (-qbz_{\rm c};q)_\infty} {(-d/z_{\rm c};q)_\infty (-dz_{\rm c};q)_\infty (-z_{\rm c}/b;q)_\infty (-qbz;q)_\infty}\right] \nonumber\\
&&{} +\lambda_1 \ln \left[ \frac{(qz_{\rm c}/z;q)_\infty (qzz_{\rm
      c};q)_\infty^2}{(q z/z_{\rm c};q)_\infty (qz_{\rm
      c}^2;q)_\infty^2 }\right] + \lambda_1 \ln\left(\frac{z-z_{\rm
    c}^{-1}}{z_{\rm c}-z_{\rm c}^{-1}}\right),
\label{eq:y1sol1}
\eea
\bea
y_2(z) &=& \kappa_2 -\lambda_2 \ln(-z_{\rm c}) -\frac{\mu_2}{z_{\rm
    c}} \nonumber\\
&&{}+\mu_2\left[\psi_1(z|q^{-1})-\psi_1(z|q)+2\psi_1(z^{-1}|q^{-1})\right]
\nonumber\\ 
&& {} +\lambda_2 \ln \left[ \frac{(qz_{\rm c}/z;q)_\infty (qzz_{\rm c};q)_\infty^2}{(q z/z_{\rm c};q)_\infty(qz_{\rm c}^2;q)_\infty^2 }\right] + \lambda_2 \ln\left(\frac{z-z_{\rm c}^{-1}}{z_{\rm c}-z_{\rm c}^{-1}}\right) \nonumber\\
&&{}+ \frac{\mu_2}{z_{\rm c}^2} \left(\frac{1}{z-z_{\rm c}^{-1}} -
\frac{1}{z_{\rm c}-z_{\rm c}^{-1}}\right), \label{eq:y2sol}
\eea
\bea
y_3(z) &=& \kappa_3 -\lambda_3 \ln(-z_{\rm c}) -\frac{\mu_3}{z_{\rm c}} + \frac{\nu_3}{z_{\rm c}^2}\nonumber\\
&&{}+\mu_3\left[\psi_1(z|q^{-1})-\psi_1(z|q)+2\psi_1(z^{-1}|q^{-1})\right]
\nonumber\\
&&{}+\nu_3\left[\psi_2(z|q)-\psi_2(z|q^{-1})-2\psi_2(z^{-1}|q^{-1})\right]
\nonumber\\
&& {} +\lambda_3 \ln \left[ \frac{(qz_{\rm c}/z;q)_\infty (qzz_{\rm c};q)_\infty^2}{(q z/z_{\rm c};q)_\infty(qz_{\rm c}^2;q)_\infty^2 }\right] + \lambda_3 \ln\left(\frac{z-z_{\rm c}^{-1}}{z_{\rm c}-z_{\rm c}^{-1}}\right)\nonumber\\
&&{}+ \left(\frac{\mu_3}{z_{\rm c}^2} -  \frac{\nu_3}{z_{\rm c}^3}\right) \left(\frac{1}{z-z_{\rm c}^{-1}} - \frac{1}{z_{\rm c}-z_{\rm c}^{-1}}\right)\nonumber\\
&&{}- \frac{\nu_3}{z_{\rm c}^3} \left(\frac{z}{\left(z-z_{\rm c}^{-1}\right)^2} - \frac{z_{\rm c}}{\left(z_{\rm c}-z_{\rm c}^{-1}\right)^2}\right).
\label{eq:y3sol}
\eea
Here $(a;q)_\infty$ denotes the q-Pochhammer symbol 
\be
(a;q)_\infty = \prod_{k=0}^{\infty} (1-aq^k),
\label{pochh}
\ee
and we have defined functions
\be
\psi_k(z|q) = \sum_{n=0}^{\infty} \frac{1}{(z_c-q^{n+1}z)^k}-
\frac{1}{z_c^k(1-q^{n+1})^k}\ .
\ee
Imposing the boundary conditions \fr{eq:xidef}, \fr{eq:yexpand} and
using the expressions presented in \ref{ap:Mcoef} for the various
constants we obtain
\be
\lambda_1=2\i,\qquad z_{\rm c}=-\frac{1}{\sqrt{ab}},
\ee
and
\be
\renewcommand{\arraystretch}{1.2}
\begin{array}{r@{\hspace{2pt}}c@{\hspace{3pt}}l}
\lambda_3 &=& \mu_2 = \lambda_2 = \kappa_1=0,\\
\nu_3 &=& \dps z_{\rm c} \mu_3 =  z_{\rm c} (z_{\rm c}-1)^2 \kappa_2 =
-\i\pi^2 \frac{z_{\rm c}^2(1-z_{\rm c})^2}{(1+z_{\rm c})^2}.
\end{array}
\ee
This is in agreement with our previous assertion \fr{eq:constants}. 
Given our result for the counting function we may then determine the
corresponding eigenvalue of the transition matrix from \fr{eq:energyMI}
\bea
\mathcal{E}_1 &=& -(1-q)\left(\frac{1}{1+a} + \frac{1}{1+b} +
\frac{2z_{\rm c}}{1-z_{\rm c}} \right)\nonumber\\
&& + \frac{1}{L^2} \frac{(1-q)\, \pi^2}{(z_{\rm c}^{-1}-z_{\rm c})}
+{\cal O}(L^{-3}),
\label{eq:MI_E}
\eea
which is the result given in \fr{E_I}.

\subsection{Region II: $-b$ inside the contour}
We now consider the case where $-b$ moves inside the contour, but
where the pole at $-1/q a$ remains outside, i.e. 
\be
b<b_{\rm c}=a^{-1/3}\quad (a>1).
\ee
The case where $a<a_{\rm c}$ is readily obtained from the results
below by the interchange $a\leftrightarrow b$. The main difference
compared to Region I is that the driving term $g_1(z)$ acquires an
additional branch point inside the contour of integration. The ${\cal
  O}(L^{-1})$ contribution $y_1(z)$ to the counting function
must therefore be determined on the basis of a different analytic
structure of the driving term $g_1(z)$:  
\bea
g_1(z) &=& -\i\ln z -\i \ln \left(\frac{z+c}{z}\right)-\i \ln
\left(\frac{z+d}{z}\right)\nonumber\\ 
&& {} +\lambda_1 \ln \left(\frac{z-qz_{\rm c}}{z}\right) -\i\ln(z+b)
+g_1^{\rm a}(z), 
\eea
where $g_1^{\rm a}(z)$ is analytic inside the contour. The solution of
the corresponding integro-differential equation proceeds along the
same lines as before, resulting in the expression \fr{eq:y1solII}
for $y_1(z)$. The solutions of the equations for $y_2(z)$ and $y_3(z)$
remain unchanged. Imposing the boundary conditions \fr{eq:xidef} imposes
$\lambda_1=3\i$ and $z_{\rm c}=-a^{-1/3}$, resulting in the
eigenvalue \fr{E_II}.  

\subsection{Region III: $-1/qa$ inside the contour}
The next case we consider is when $-1/qa$ lies inside and $-b$ outside
the integration contour, which occurs in the parameter regime $q^{1/2}
< b < q^2 a$. The driving term $g_1(z)$ of the integro-differential
equation for $y_z(z)$ is expressed as
\bea
g_1(z) &=&  -\i \ln \left(\frac{z+c}{z}\right)-\i \ln
\left(\frac{z+d}{z}\right)\nonumber\\ 
&& {} +\lambda_1 \ln \left(\frac{z-qz_{\rm c}}{z}\right) +
\i\ln\left(\frac{z+1/qa}{z}\right) +g_1^{\rm a}(z), 
\eea
where $g_1^{\rm a}(z)$ is analytic inside the contour of
integration. Proceeding as before we arrive at the result for $y_1(z)$
given in \fr{eq:y1solIII}. The results for $y_2(z)$ and $y_3(z)$ are
the same as before and are given in \fr{eq:y2sol} and \fr{eq:y3sol}.
Imposing the boundary conditions \fr{eq:xidef}
fixes $\lambda_1=\i$ and $z_{\rm c}=-q/b$, leading to the eigenvalue
given in \fr{E_III}.  

\subsection{Region IV: $-1/qa$ and $-b$ inside the contour}
The last case we consider is when both $-1/qa$ and $-b$ lie inside the
contour of integration. This occurs when $a> q^{-3/2}$ and $b<q^{1/2}$.
We may express $g_1(z)$ in the form
\bea
g_1(z) &=& -\i \ln \left(\frac{z+c}{z}\right)-\i \ln
\left(\frac{z+d}{z}\right) -\i\ln(z+b)\nonumber\\ 
&& {} +\lambda_1 \ln \left(\frac{z-qz_{\rm c}}{z}\right)  +
\i\ln\left(\frac{z+1/qa}{z}\right) +g_1^{\rm a}(z), 
\eea
where $g_1^{\rm a}(z)$ is again analytic inside the contour and then proceed
as in the other cases. Solving the integro-differential equation for
$y_1(z)$ results in \fr{eq:y1solIV}. The results for $y_2(z)$ and
$y_3(z)$ are again given by \fr{eq:y2sol} and \fr{eq:y3sol}
respectively. Imposing the boundary conditions
\fr{eq:xidef} now gives $\lambda_1=2\i$ and $z_{\rm c}=-q^{1/2}$ and
leads to the eigenvalue given in \fr{E_IV}. 
\section{Dependence on $c$ and $d$}
\label{se:cd}
\begin{figure}[ht]
\centerline{
\begin{picture}(150,100)
\put(0,0){\epsfig{width=150pt,file=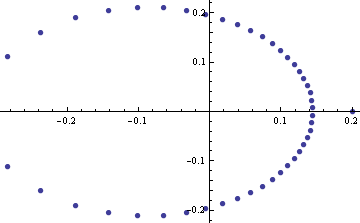}}
\end{picture}}
\caption{Root distributions for $a=\kappa^+_{\alpha,\gamma}=5$, $b=\kappa^+_{\beta,\delta}=2$, $c=\kappa^-_{\alpha,\gamma}=-0.2$, $d=\kappa^-_{\beta,\delta}=-0.023$, $q=0.1$ and $L=50$}. 
\label{fig:roots2}
\end{figure}

In all calculations described above we have assumed that both $-c$ and
$-d$ lie inside the contour of integration. This is the case if
\be
-c < z^* \quad\textrm{and}\quad -d < z^*,
\ee
where $z^*$ is the point where the contour crosses the positive real
axis, i.e. the solution of the equation $Y_L(z^*)=0$. In leading order
this is determined by the solution of 
\be
-\frac{z^*}{z_{\rm c}} \frac{(1-z_{\rm c})^2}{(1-z^*)^2}=1.
\ee
This condition is easily solved for $z^*$ as a function of $z_c$ and
using the explicit expressions for $z_{\rm c}$ in Regions I-IV we obtain
corresponding restrictions on the allowed values of $c$ and
$d$. For example, in Region I we find
\be
-c,-d< z^*(a,b)=\frac{1+4\sqrt{ab}+ab-\sqrt{(1+4\sqrt{ab}+ab)^2-4ab}}{2\sqrt{ab}}.
\label{eq:rangecd}
\ee
However, as we will now show, the results in the high and low density
phases presented above have a somewhat larger realm of validity than
suggested by \fr{eq:rangecd}. To that end let us consider the
situation where $-d$ is still inside the contour of integration, but
$-c$ is slightly larger than $z^*(a,b)$. Then the root distribution
corresponding to the largest eigenvalue has the same set of integers
as before 
\be
Y_L(z_j) = -\pi +\frac{2\pi j}{L}, 
\ee
but the root distribution now has an isolated root $\zeta$ lying
outside the contour of integration. The position of the isolated root
is 
\be
\zeta=-c+{\cal O}(e^{-\nu L}),
\ee
where $\nu>0$. In Fig.~\ref{fig:roots2} the root distribution
is depicted for a specific case with $c=-0.2$, and it can be seen that
$\zeta\approx -c$. The value of $z^*(a,b)$ in this case is
approximately $0.136$.

In order to turn the summation over this distribution of roots into an
integral, the contour $C_2$ of Figure~\ref{fig:contour} has to be
extended to include the isolated root $\zeta$. This is achieved by adding
a contour that runs from $z^*+\i0$ to $\zeta+\i0$ and back from
$\zeta-\i0$ to $z^*-\i0$. Importantly the extra contour does not
encircle any other poles or branch points of the counting function.
Hence the analysis of this case is exactly the same as before, and the
eigenvalue is again given by \fr{eq:MI_E}  

The situation becomes more complicated when $-c$ is increased
further. Now several isolated roots may lie on the positive real axis
outside the contour, close to points $-q^n c$ for $n\in\mathbb{N}$. Naively the same argument as for the case of a
single isolated roots applies, but a detailed analysis of such cases
is beyond the scope of this publication.
\section{Conclusions}
\label{se:summ}
In this work we have analyzed the Bethe ansatz equations of the
partially asymmetric exclusion process with open boundaries. We have
focussed on the parameter regime corresponding to low and high density
phases in the stationary state. We have determined the eigenvalue of
the transition matrix with the largest non-zero real part, which
characterizes the relaxation towards the stationary state at
asymptotically late times. We found that both the low and high
density phases are subdivided into several regimes, which are
characterized by different relaxational behaviours. In the vicinity of
the coexistence line which separates low and high density phases the
relaxational behaviour can be understood in terms of diffusion of
domain walls. In the other regimes such interpretations are still
possible, but are not conclusively supported by the available results.
A number of open questions remain. We have not studied the parameter
regime corresponding to the maximal current phase at stationarity.
Furthermore, our analysis has been restricted to the ``forward bias
regime'', in which the injection/extraction rates at the boundaries
are compatible with the bias present in the bulk. It would be
interesting to extend our analysis to the maximum current phase as
well as the ``reverse bias regime''. Another open issue
is the precise physical nature of the relaxational mechanism
sufficiently far away from the coexistence line. Is the relaxation
still driven by diffusion of some kind of domain walls, or does
another mechanism take over? This issue is particularly relevant in
Region III, which occupies a large part of the phase diagram when
$q\rightarrow 1$. For symmetric diffusion $q=1$ the relaxation is
known to be diffusive and the spectral gap scales as
$\mathcal{O}(L^{-2})$. Further interesting questions are whether the
Bethe ansatz solution of the open XXZ chain can be used to calculate
current fluctuations \cite{currentfluc} and whether it is possible to
determine correlation functions by means of Bethe ansatz
\cite{maillet}. Perhaps recent algebraic insights may offer new tools 
for further analysis. Such insights include the PASEP being 
an exceptional representation of the two-boundary Temperley-Lieb algebra 
\cite{GierN} through the connection with the open spin chain \cite{GierE06}, 
or the connection with tridiagonal algebras \cite{baseil,aneva}.

\ack
We are grateful to Gunter Sch\"utz and Robin Stinchcombe for helpful
discussions. This work was supported by the ARC (JdG), the EPSRC under
grant EP/D050952/1 (FE) and the John Fell OUP Research Fund.

\appendix
\section{Details on the calculation of the counting function} 
\label{ap:countingfunction}
In this Appendix we describe how to solve the integro-differential
equation for the counting function. We start by deriving a simple
identity that proves to be very useful in the subsequent analysis. Let
$C$ be the contour of integration from  $z_{\rm c}^-$ to $z_{\rm c}^+$
defined by the locus of roots, c.f. Fig.\ref{fig:contour}. Let $D$
denote the interior of $C$ and $f'(z)$ be an analytic function in $D$.
Then elementary considerations show that
\renewcommand{\arraystretch}{1.5}
\bea
\frac{1}{2\pi\i} \int_{z_{\rm c}^-}^{z_{\rm c}^+} \ln (w-z)\left(\frac{1}{w-a} + f'(w)\right)
\,\d w &=& \nonumber\\
&&\hphantom{\hspace{-6cm}} \left\{ 
\begin{array}{ll}
\displaystyle\ln\left(\frac{z_{\rm c}-a}{z-a}\right) +f(z_{\rm c})-f(z) \quad & {\rm if}\ a\not\in D,\ z \in D,\\
\ln(a-z_{\rm c}) + f(z_{\rm c})-f(z)&  {\rm if}\ a,z\in D.
\end{array}\right.
\label{eq:logint}
\eea
Here, for every $w$, we have placed the branch cut of $\ln (w-z)$ along the line from $z$ to $z_{\rm c}$ so that the left hand side  is a well defined contour integral. The identity \r{eq:logint} is useful for constructing solutions of the
integro-differential equations \fr{eq:yn}. In particular, if $y(z)$ is
given by
\be
y(z) = y^{\rm a}(z) + \sum_m A_m \ln(z-z_m),
\ee
where $\d y^{\rm a}(z)/\d z$ is analytic in $D$, then, for $z$ close to the locus of the roots,
\be
\frac{1}{2\pi\i} \oint_{z_{\rm c}} K(w,z) y'(w) \d w = y(qz) -
y(z_{\rm c}) + \sum_m A_m K(z_m,z). 
\label{eq:logintK}
\ee
Here the kernel $K(w,z)$ is given in \fr{eq:KernelDef} whose branch
cut we take as above. The condition on $z$ is that $1/z$, $1/q^2z$ and
$z/q$ all lie outside the contour of integration ($z/q$ lies outside
the contour as we take $z$ to be very close to the latter).
\subsection{Region I: large values of $a$ and $b$}
%
We first consider the case
\begin{enumerate}
\item{} $-c$ and $-d$ are inside $D$,
\item{} $\pm 1$, $-a$, $-b$, $-1/qa$ and $-1/qb$ are outside $D$.
\end{enumerate}
Loosely speaking this corresponds to small $q$, large positive $a$ and
$b$ and small negative $c$ and $d$. We further assume (and verify a
posteriori) that $D$ lies inside the unit circle, and hence that the
point $1/qc$ and $1/qd$ also lie outside $D$. 

We wish to solve the system of integro-differential equations \fr{eq:yn}:
\be
y_n(z) = g_n(z) + \frac{1}{2\pi \i} \int_{z_{\rm c}^-}^{z_{\rm c}^+}
K(w,z) y'_n(w) \,\d w,
\label{eq:yn2}
\ee
where the driving terms $g_n(z)$ are given in \fr{eq:gs}. We note that
the equations are coupled as the driving terms of the equations for
larger $n$ depend on the solutions for smaller values of $n$. In the
following we construct a solution of \fr{eq:yn2} and verify a
posteriori that the above assumptions hold.

\subsubsection{Equation for $y_0(z)$:}
\paragraph{}

The driving term of the leading equation is given by
\be
g_0(z) = -\i g(z) = -\i\ln z -2\i \ln\left(\frac{1-qz}{1-z}\right) =
-\i\ln z + g_0^{\rm a}(z).
\ee
Here $g_0^a(z)$ is analytic in $D$. We assume that $y_0(z)$ has the
same analytic structure in $D$, i.e.
\be
y_0(z) = -\i \ln z + y_0^{\rm a}(z),
\label{eq:y0}
\ee
where $y_0^{\rm a}(z)$ is analytic. Subsituting this ansatz into the
integro-differential equation \fr{eq:yn2} for $n=0$ we obtain from \r{eq:logintK},
\be
y_0^{\rm a}(z) = g_0^{\rm a}(z) -\i K(0,z) + y_0(qz) - y_0(z_{\rm c}).
\label{eq:y0a}
\ee
Combining equations \fr{eq:y0} and \fr{eq:y0a} we obtain a functional
equation for $y_0^{\rm a}(z)$
\be
y_0^{\rm a}(z) - y_0^{\rm a}(qz) = -2\i
\ln\left(\frac{1-qz}{1-z}\right) + \i \ln(-z_{\rm c}) - y_0^{\rm
  a}(z_{\rm c}).
\ee
The constant $y_0^{\rm a}(z_{\rm c})$ is readily determined by setting
$z=0$. The resulting functional equation is then readily solved, giving 
\be
y_0^{\rm a}(z) = 2\i\ln(1-z) -\i \ln\left( -\frac{(1-z_{\rm
    c})^2}{z_{\rm c}}\right). 
\ee
The zeroeth order term in the expansion of the counting function is
then found to be
\be
y_{0}(z) = -\i \ln \left[ - \frac{z}{z_{\rm c}}
\left(\frac{1-z_{\rm c}}{1-z}\right)^2 \right].
\ee

\subsubsection{Equation for $y_1(z)$:}
\paragraph{}

With the assumptions on $q,a,b,c$ and $d$ stated above the driving term
\fr{eq:gs} of the integro-differential equation \fr{eq:yn2} for $n=1$
can be cast in the form
\bea
g_1(z) &=& -\i\ln z -\i \ln \left(\frac{z+c}{z}\right)-\i \ln
\left(\frac{z+d}{z}\right)\nonumber\\ 
&& {} +\lambda_1 \ln \left(\frac{z-qz_{\rm c}}{z}\right) +g_1^{\rm a}(z),
\eea
where $g_1^{\rm a}(z)$ is analytic in $D$. The singularities of
$y_1(z)$ inside the domain $D$ can be inferred by trying to solve the
integro-differential equation by iteration. It is then quickly seen
that a branch point in $y_1(z)$ at $z=-c$ produces a branch point at
$z=-qc$, which in turn leads to a branch point at $z=-q^2c$ {\sl etc}.
This suggests the following ansatz for $y_1(z)$ 
\bea
y_1(z) &=& -\i \ln \left[ (-c/z;q)_\infty (-d/z;q)_\infty \right] +\lambda_1 \ln (qz_{\rm c}/z;q)_\infty \nonumber\\
&& {} -\i\ln z +y_1^{\rm a}(z),
\label{eq:y1ansatz}
\eea
where $y_1^{\rm a}(z)$ is analytic inside $D$, and the q-Pochhammer
symbol $(a;q)_\infty$ was defined in \fr{pochh}. Substituting
\fr{eq:y1ansatz} into \fr{eq:yn2} we find 
\bea
y_1^{\rm a}(z) &=& \i \ln \left[ (-c/qz;q)_\infty (-d/qz;q)_\infty \right] -\lambda_1 \ln (z_{\rm c}/z;q)_\infty \nonumber\\
&&{} -\i\ln \left[ (1+cz)(1+qcz) (1+dz)(1+qdz)\right] \nonumber\\
&&{} +\lambda_1\ln \left[ (1-qz_{\rm c}z)(1-q^2z_{\rm c}z)\right]+g_1^{\rm a}(z) - y_1(z_{\rm c})\nonumber\\
&&{} +\i\ln (-qz)+ y_1(qz).
\label{eq:y1analytic}
\eea
This leads to the following functional equation for $y_1^{\rm a}(z)$
\bea
y_1^{\rm a}(z) &=& -\i\ln \left[ (1+cz)(1+qcz) (1+dz)(1+qdz)\right] \nonumber\\
&&{} +\lambda_1\ln \left[ (1-qz_{\rm c}z)(1-q^2z_{\rm c}z)\right]\nonumber\\
&&{} +g_1^{\rm a}(z) - y_1(z_{\rm c}) \pm\pi+y_1^{\rm a}(qz).
\label{eq:y1analytic2}
\eea
The value of $y_1(z_{\rm c})$ is easily determined by evaluating
\fr{eq:y1analytic2} at $z=0$
\be
y_1(z_{\rm c}) = \pm\pi + g_1^{\rm a}(0).
\ee
Using the fact that for two analytic functions $u$ and $v$, the equation 
\be
u(z)-u(qz)=v(z),
\ee
is solved by $u(z)=\sum_{k=0}^\infty v(q^kz)$, it is now a straightforward matter to solve the functional equation
\fr{eq:y1analytic2}, leading to the result given in equation \fr{eq:y1sol1}.
%

\subsubsection{Equation for $y_2(z)$:}
\paragraph{}
The driving term \fr{eq:gs} of the integro-differential equation
\fr{eq:yn} for $n=2$ can be represented in the form
\be
g_2(z) = \lambda_2 \ln \left(\frac{z-qz_{\rm c}}{z}\right) -\mu_2
\frac{q}{z-qz_{\rm c}} + g_2^{\rm a}(z), 
\ee
where $g_2^{\rm a}(z)$ is analytic in $D$. As in the case $n=1$ the
singularities of $y_2(z)$ inside $D$ can be determined by attempting
to solve the equation by iteration. This results in the ansatz
\bea
y_2(z) &=& \lambda_2 \ln (qz_{\rm c}/z;q)_\infty -
\mu_2\sum_{n=0}^\infty \frac{q^{n+1}}{z-q^{n+1} z_{\rm c}} + y_2^{\rm
  a}(z),
\label{eq:y2ansatz}
\eea
where $y_2^{\rm a}(z)$ is analytic inside $D$. Substituting
\fr{eq:y2ansatz} into \fr{eq:yn2} we obtain a functional equation for
$y_2^{\rm a}(z)$
\bea
y_2^{\rm a}(z) -y_2^{\rm a}(qz) &=& 
\kappa_2-\lambda_2\ln(-z_c)
+\lambda_2\ln \left[\frac{(1-z_{\rm c}z)(1-qz_{\rm c}z)}{1-qz/z_{\rm
      c}}\right]\nonumber\\
&-&\mu_2 \left[\frac{qz}{1-qz_{\rm c}z} + 
\frac{z}{1-z_{\rm    c}z} -\frac{1}{qz-z_{\rm    c}} \right]-y_2(z_c).
\label{eq:y2analytic}
\eea
Evaluating \fr{eq:y2analytic} at $z=0$ fixes the constant to be
\be
y_2(z_{\rm c}) =\kappa_2 -\lambda_2 \ln(-z_{\rm c}) -\frac{\mu_2}{z_{\rm c}}.
\ee
The functional equation \fr{eq:y2analytic} is then solved by
elementary means, resulting in the expression for $y_2(z)$ given in
equation \fr{eq:y2sol}.

\subsubsection{Equation for $y_3(z)$:}
\paragraph{}
The driving term \fr{eq:gs} of the integro-differential equation
\fr{eq:yn} for $n=3$ can be represented in the form
\be
g_3(z) = \lambda_3 \ln \left(\frac{z-qz_{\rm c}}{z}\right) -\mu_3 \frac{q}{z-qz_{\rm c}} -\nu_3 \frac{q^2}{(z-qz_{\rm c})^2} + g_3^{\rm a}(z),
\ee
where $g_3^{\rm a}(z)$ is analytic in $D$. Determining the
singularities of $y_3(z)$ inside $D$ by iterating the
integro-differential equation now results in the ansatz
\bea
y_3(z) &=& \lambda_3 \ln (qz_{\rm c}/z;q)_\infty -
\mu_3\sum_{n=0}^\infty \frac{q^{n+1}}{z-q^{n+1} z_{\rm c}}\nonumber\\ 
&&{}-\nu_3\sum_{n=0}^\infty \frac{q^{2n+2}}{(z-q^{n+1} z_{\rm c})^2}
+ y_3^{\rm a}(z),
\label{eq:y3ansatz}
\eea
where $y_3^{\rm a}(z)$ is analytic inside $D$. Substituting
\fr{eq:y3ansatz} into \fr{eq:yn2} for $n=3$ gives a functional
equation for $y_3^{\rm a}(z)$
\bea
y_3^{\rm a}(z) -y_3^{\rm a}(qz) &=& 
\kappa_3-\lambda_3\ln(-z_c)-y_3(z_c)\nonumber\\
&+&\lambda_3\ln \left[\frac{(1-z_{\rm c}z)(1-qz_{\rm c}z)}{1-qz/z_{\rm
      c}}\right]\nonumber\\
&-&\mu_3 \left[ \frac{qz}{1-qz_{\rm c}z} + 
\frac{z}{1-z_{\rm    c}z} -\frac{1}{qz-z_{\rm    c}}
\right]\nonumber\\
&-&\nu_3 \left[ \frac{q^2z^2}{(1-qz_{\rm c}z)^2} + 
\frac{z^2}{(1-z_{\rm    c}z)^2} -\frac{1}{(qz-z_{\rm    c})^2}
\right].
\label{eq:y3analytic}
\eea
Evaluating \fr{eq:y3analytic} at $z=0$ again fixed the constant
\be
y_3(z_{\rm c}) = \kappa_3 -\lambda_3 \ln(-z_{\rm c}) -\frac{\mu_3}{z_{\rm c}} + \frac{\nu_3}{z_{\rm c}^2}.
\ee
Solving the functional equation \fr{eq:y3analytic} then results
in the expression for $y_3(z)$ given in equation \fr{eq:y3sol}.
%
\subsection{Region II: $-b$ inside the contour of integration}
The determination of $y_n(z)$ for $n\geq 2$ is exactly the same as in
the previous section and in particular the expressions for
$y_2(z)$ and $y_3(z)$ are unchanged and given by \fr{eq:y2sol} and \fr{eq:y3sol}.

\subsubsection{Equation for $y_1(z)$:}
\paragraph{}
With $-b$ lying inside $D$ the driving term $g_1(z)$ may be expressed as
\bea
g_1(z) &=& -2\i\ln z -\i \ln \left(\frac{z+c}{z}\right)-\i \ln
\left(\frac{z+d}{z}\right)\nonumber\\ 
&& {} +\lambda_1 \ln \left(\frac{z-qz_{\rm c}}{z}\right) -\i \ln
\left(\frac{z+b}{z}\right) +g_{1,{\rm II}}^{\rm a}(z), 
\eea
where $g_{1,{\rm II}}^{\rm a}(z)$ is analytic in $D$. Proceeding as
before, we arrive at the following ansatz for $y_1(z)$
\bea
y_1(z) &=& -\i \ln \left[ (-c/z;q)_\infty (-d/z;q)_\infty \right] +\lambda_1 \ln (qz_{\rm c}/z;q)_\infty \nonumber\\
&& {} - \i \ln( -b/z;q)_\infty -2\i\ln z +y_1^{\rm a}(z),
\label{eq:y1ansatzII}
\eea
where $y_1^{\rm a}(z)$ is analytic in $D$. Substituting
\fr{eq:y1ansatzII} into \fr{eq:yn2} we obtain the functional equation
\bea
y_1^{\rm a}(z) &=& -\i\ln \left[ (1+cz)(1+qcz) (1+dz)(1+qdz)\right] \nonumber\\
&&{} +\lambda_1\ln \left[ (1-qz_{\rm c}z)(1-q^2z_{\rm c}z)\right] - y_1(z_{\rm c}) \pm 2\pi \nonumber\\
&&{}-\i\ln \left[ (1+bz)(1+qbz)\right]+g_{1,{\rm II}}^{\rm a}(z) +y_1^{\rm a}(qz).
\label{eq:y1analytic2II}
\eea
Evaluating \fr{eq:y1analytic2II} at $z=0$ again fixes the constant
\be
y_1(z_{\rm c}) = \pm 2\pi + g_{1,{\rm II}}^{\rm a}(0),
\ee
and solving \fr{eq:y1analytic2II} then results in the following expression for
$y_1(z)$ 
\bea
y_1(z) &=& -2\i\ln\left[ -\frac{z}{z_{\rm c}}\right]-\i\ln\left[ \frac{1-z_{\rm c}^2}{1-z^2}\right] +\kappa_1 - \i\ln\left(a\right) -\lambda_1\ln(-z_{\rm c}) \nonumber\\
&&{} -\i\ln\left[\frac{(-c/z;q)_\infty (-cz;q)_\infty (-z/a;q)_\infty (-qaz_{\rm c};q)_\infty} {(-c/z_{\rm c};q)_\infty (-cz_{\rm c};q)_\infty (-z_{\rm c}/a;q)_\infty (-qaz;q)_\infty}\right] \nonumber\\
&&{}-\i\ln\left[\frac{(-d/z;q)_\infty (-dz;q)_\infty (-b/z;q)_\infty (-bz;q)_\infty } {(-d/z_{\rm c};q)_\infty (-dz_{\rm c};q)_\infty (-b/z_{\rm c};q)_\infty (-bz_{\rm c};q)_\infty }\right] \nonumber\\
&&{} +\lambda_1 \ln \left[ \frac{(qz_{\rm c}/z;q)_\infty (qzz_{\rm c};q)_\infty^2}{(q z/z_{\rm c};q)_\infty (qz_{\rm c}^2;q)_\infty^2 }\right] + \lambda_1 \ln\left(\frac{z-z_{\rm c}^{-1}}{z_{\rm c}-z_{\rm c}^{-1}}\right).
\label{eq:y1solII}
\eea
%

\subsection{$-1/qa$ inside the contour of integration}
The determination of $y_n(z)$ for $n\geq 2$ is exactly the same as before
and in particular the expressions for
$y_2(z)$ and $y_3(z)$ are unchanged and given by \fr{eq:y2sol} and \fr{eq:y3sol}.

\subsubsection{Equation for $y_1(z)$:}
\paragraph{}

In this parameter regime  $g_1(z)$ is expressed as
\bea
g_1(z) &=& -\i \ln \left(\frac{z+c}{z}\right)-\i \ln
\left(\frac{z+d}{z}\right)\nonumber\\ 
&& {} +\lambda_1 \ln \left(\frac{z-qz_{\rm c}}{z}\right) +\i \ln
\left(\frac{z+1/qa}{z}\right) +g_{1,{\rm III}}^{\rm a}(z), 
\eea
where $g_{1,{\rm III}}^{\rm a}(z)$ is analytic inside $D$. The
appropriate ansatz for $y_1(z)$ takes the form
\bea
y_1(z) &=& -\i \ln \left[ (-c/z;q)_\infty (-d/z;q)_\infty \right]
+\lambda_1 \ln (qz_{\rm c}/z;q)_\infty \nonumber\\ 
&& {} + \i \ln( -1/qaz;q)_\infty +y_1^{\rm a}(z),
\label{eq:y1ansatzIII}
\eea
where $y_1^{\rm a}(z)$ is analytic inside $D$. 
Substituting \fr{eq:y1ansatzIII} into \fr{eq:yn2} we arrive at the
functional equation
\bea
y_1^{\rm a}(z) &=& -\i\ln \left[ (1+cz)(1+qcz) (1+dz)(1+qdz)\right] \nonumber\\
&&{} +\lambda_1\ln \left[ (1-qz_{\rm c}z)(1-q^2z_{\rm c}z)\right] - y_1(z_{\rm c}) \nonumber\\
&&{}+ \i\ln \left[ (1+z/qa)(1+z/a)\right]+g_{1,{\rm III}}^{\rm a}(z) +y_1^{\rm a}(qz).
\label{eq:y1analytic2III}
\eea
Evaluating \fr{eq:y1analytic2III} at $z=0$ gives
\be
y_1(z_{\rm c}) = g_{1,{\rm III}}^{\rm a}(0),
\ee
and we finally arrive at the following expression for $y_1(z)$
\bea
y_1(z) &=& -\i\ln\left[ \frac{1-z_{\rm c}^2}{1-z^2}\right] +\kappa_1
- \i\ln\left(b/q\right) -\lambda_1\ln(-z_{\rm c}) \nonumber\\ 
&&{} -\i\ln\left[\frac{(-c/z;q)_\infty (-cz;q)_\infty (-1/qaz_{\rm
c};q)_\infty(-z_c/qa;q)_\infty} {(-c/z_{\rm c};q)_\infty (-cz_{\rm
c};q)_\infty 
(-1/qaz;q)_\infty(-z/qa;q)_\infty}\right] \nonumber\\ 
&&{}-\i\ln\left[\frac{(-d/z;q)_\infty (-dz;q)_\infty (-z/b;q)_\infty
    (-qbz_{\rm c};q)_\infty} {(-d/z_{\rm c};q)_\infty (-dz_{\rm
      c};q)_\infty (-z_{\rm c}/b;q)_\infty (-qbz;q)_\infty}\right]
\nonumber\\ 
&&{} +\lambda_1 \ln \left[ \frac{(qz_{\rm c}/z;q)_\infty (qzz_{\rm
      c};q)_\infty^2}{(q z/z_{\rm c};q)_\infty (qz_{\rm
      c}^2;q)_\infty^2 }\right] 
+\lambda_1
\ln\left(\frac{z-z_{\rm c}^{-1}}{z_{\rm c}-z_{\rm c}^{-1}}\right). 
\label{eq:y1solIII}
\eea
%
\subsection{$-b$ and $-1/qa$ inside the contour of integration}
The determination of $y_n(z)$ for $n\geq 2$ is exactly the same as before
and in particular the expressions for $y_2(z)$ and $y_3(z)$ are
unchanged and given by \fr{eq:y2sol} and \fr{eq:y3sol}. 

\subsubsection{Equation for $y_1(z)$:}
\paragraph{}
Now the driving term is expressed as
\bea
g_1(z) &=& -\i\ln z -\i \ln \left(\frac{z+c}{z}\right)-\i \ln
\left(\frac{z+d}{z}\right) -\i\ln\left(\frac{z+b}{z}\right)\nonumber\\ 
&& {} +\lambda_1 \ln \left(\frac{z-qz_{\rm c}}{z}\right) +\i \ln
\left(\frac{z+1/qa}{z}\right) +g_{1,{\rm IV}}^{\rm a}(z), 
\eea
where $g_{1,{\rm IV}}^{\rm a}(z)$ is analytic in $D$. Our ansatz for
$y_1(z)$ then takes the form 
\bea
y_1(z) &=& -\i\ln z -\i \ln \left[ (-c/z;q)_\infty (-d/z;q)_\infty \right] +\lambda_1 \ln (qz_{\rm c}/z;q)_\infty \nonumber\\
&& {} - \i \ln( -b/z;q)_\infty + \i \ln( -1/qaz;q)_\infty +y_1^{\rm a}(z),
\label{eq:y1ansatzIV}
\eea
where $y_1^{\rm a}(z)$ is analytic. Substituting \fr{eq:y1ansatzIV}
into \fr{eq:yn2} we obtain the functional equation
\bea
y_1^{\rm a}(z) &=& -\i\ln \left[ (1+cz)(1+qcz) (1+dz)(1+qdz)\right] \nonumber\\
&&{} +\lambda_1\ln \left[ (1-qz_{\rm c}z)(1-q^2z_{\rm c}z)\right] - y_1(z_{\rm c}) \pm \pi \nonumber\\
&&{}+ \i\ln \left[ (1+z/qa)(1+z/a)\right] - \i\ln \left[ (1+bz)(1+qbz)\right] \nonumber\\
&&{} +g_{1,{\rm IV}}^{\rm a}(z) +y_1^{\rm a}(qz).
\label{eq:y1analytic2IV}
\eea
Fixing the constant as before gives
\be
y_1(z_{\rm c}) = \pm\pi + g_{1,{\rm IV}}^{\rm a}(0),
\ee
resulting in 
\bea
y_1(z) &=& -\i \ln\left[ -\frac{z}{z_{\rm c}}\right] -\i\ln\left[
  \frac{1-z_{\rm c}^2}{1-z^2}\right] +\kappa_1 - \i\ln\left(q\right)
-\lambda_1\ln(-z_{\rm c}) \nonumber\\ 
&&{} -\i\ln\left[\frac{(-c/z;q)_\infty (-cz;q)_\infty (-1/qaz_{\rm
      c};q)_\infty(-z_{\rm c}/qa;q)_\infty} {(-c/z_{\rm c};q)_\infty
    (-cz_{\rm c};q)_\infty     (-1/qaz;q)_\infty(-z/qa;q)_\infty
}\right] \nonumber\\ 
&&{}-\i\ln\left[\frac{(-d/z;q)_\infty (-dz;q)_\infty (-b/z;q)_\infty
    (-bz;q)_\infty} {(-d/z_{\rm c};q)_\infty (-dz_{\rm c};q)_\infty
    (-b/z_{\rm c};q)_\infty (-bz_{\rm c};q)_\infty}\right] \nonumber\\ 
&&{} +\lambda_1 \ln \left[ \frac{(qz_{\rm c}/z;q)_\infty (qzz_{\rm
      c};q)_\infty^2}{(q z/z_{\rm c};q)_\infty (qz_{\rm
      c}^2;q)_\infty^2 }\right] 
\lambda_1
\ln\left(\frac{z-z_{\rm c}^{-1}}{z_{\rm c}-z_{\rm c}^{-1}}\right). 
\label{eq:y1solIV}
\eea
%

\section{Analysis of the Abel-Plana Formula}
\label{app:abel}
In this appendix we sketch how to extract the finite-size correction
terms from the integral expression
\bea
\i\,Y_L(z) &=& g(z) + \frac{1}{L} g_{\rm b}(z) +\frac{1}{2\pi}
\int_{\xi^*}^{\xi} K(w,z) Y'_L(w)\ \d w \nonumber \\ 
&& \hspace{-1cm}+ \frac{1}{2\pi} \int_{C_1} \frac{K(w,z)Y'_L(w)}{1-\e^{-\i L
Y_L(w)}}\ \d w + \frac{1}{2\pi} \int_{C_2} \frac{K(w,z)Y'_L(w)}{
\e^{\i L Y_L(w)}-1}\ \d w.
\label{eq:intYapp}
\eea
The main contributions to the correction terms in \fr{eq:intYapp} comes
from the vicinities of the endpoints $\xi$, $\xi^*$. Along the contour
$C_1$ the imaginary part of the counting function is positive, ${\rm
  Im}(Y_L(w))>0$, whereas along the contour $C_2$ it is negative,
${\rm Im}(Y_L(w))<0$. As a result the integrands decay exponentially
with respect to the distance from the endpoints. In the vicinity of
$\xi$, we therefore expand 
\be
Y_L(w)=Y_L(\xi)+Y_L'(\xi) (w-\xi)+ \ldots\ .
\ee
Assuming that $Y_L'(\xi)$ is ${\cal O}(1)$ (an assumption that will be
checked self-consistently), we find that the leading contribution for
large $L$ is given by
\bea
\frac{1}{2\pi}
\int_{C_1} \frac{K(w,z)Y'_L(w)}{1-\e^{-\i L Y_L(w)}}\ \d w &\sim&
\frac{Y_L'(\xi)}{2\pi} \int_{\xi}^0 \frac{K(w,z)}{1+\e^{-\i L
    Y_L'(\xi)(w-\xi)}}\ \d w\ ,\nn
&& \mbox{} - (\xi \rightarrow \xi^*).
\eea
Here we have used the boundary conditions \fr{eq:xidef}. Carrying out
the analogous analysis for the integral along the contour $C_2$, we
arrive at the following expression for the leading contribution of the
last two terms in \fr{eq:intYapp} 
\bea
{\cal A} &=& \frac{1}{2\pi} \int_{C_1} \frac{K(w,z)Y'_L(w)}{1-\e^{-\i L
Y_L(w)}}\ \d w + \frac{1}{2\pi} \int_{C_2} \frac{K(w,z)Y'_L(w)}{
\e^{\i L Y_L(w)}-1}\ \d w\nn
&=& \frac{\i}{2\pi L}\int_0^\infty \frac{1}{1+e^x}\left[
K\Bigl(\xi+\frac{\i\, x}{LY_L'(\xi)},z\Bigr)-
K\Bigl(\xi-\frac{\i\, x}{LY_L'(\xi)},z\Bigr)
\right] \d x\nn
&& \mbox{} - (\xi \rightarrow \xi^*).
\label{eq:fscorr1}
\eea
If the endpoints $\xi,\xi^*$ are such that we can Taylor-expand the
kernels appearing in \fr{eq:fscorr1}, we can simplify the expression
further with the result
\bea
{\cal A} &\approx& - \frac{K'(\xi,z)}{\pi L^2 Y_L'(\xi)}\int_0^\infty \frac{x}{1+e^x}\ \d x\ -\ (\xi \rightarrow \xi^*)\nn
&=& -\frac{\pi}{12L^2}\frac{K'(\xi,z)}{Y_L'(\xi)}\ -\ (\xi \rightarrow \xi^*).
\label{eq:fscorr2}
\eea
This is the leading Euler-MacLaurin correction term that occurs in
the low and high density phases, see \fr{eq:intY_M}. The key in the
above derivation was the ability to expand
\bea
K\Bigl(\xi+\frac{\i\, x}{LY_L'(\xi)},z\Bigr)&-&
K\Bigl(\xi-\frac{\i\, x}{LY_L'(\xi)},z\Bigr)
\nn
&&\sim
\ln\left[\frac{LY_L'(\xi)\bigl(z^{-1}-\xi\bigr)+\i\, x}
{LY_L'(\xi)\bigl(z^{-1}-\xi\bigr)-\i\, x}\right]
\eea
in a power series in $x$. This is unproblematic as long as
$LY_L'(\xi)\bigl(z^{-1}-\xi\bigr)$ is large, which turns out to be the
case in the low and high density phases as well as on the coexistence
line.

\section{Expansion coefficients\label{ap:Mcoef}}
In this appendix we list the coefficients arising in the expansion
(\ref{eq:yn}), (\ref{eq:gs}) of the integral equation for the counting
function $Y_L(z)$. In the following list we abbreviate $y'_n(z_{\rm
  c})$ by $y'_n$. We note that by definition $\delta_n$ and $\eta_n$
are real quantities.
\bea
\kappa_1 &=& - y_0' \delta_1,\\
\lambda_1 &=&  y_0' \frac{\eta_1}{\pi},\\
\kappa_2 &=& -y_0' \delta_2 - y_1' \delta_1 -
  \frac12 y_0'' (\delta_1^2 - \eta_1^2), \\ 
\lambda_2 &=& \frac1{\pi}\left( y'_0 \eta_2 + y'_1\eta_1 + y_0'' \delta_1
\eta_1\right),\\ 
\mu_2 &=& y'_0 \frac{\delta_1\eta_1}{\pi},\\
\kappa_3 &=& -y_0'\delta_3 - y_1'\delta_2 -y_2'\delta_1 - y_0'' (\delta_1\delta_2 -
\eta_1\eta_2) -\frac12 y_1''(\delta_1^2-\eta_1^2) \nonumber\\
&& {} - \frac16 y_0'''\delta_1(\delta_1^2 - 3\eta_1^2), \\
\lambda_3 &=& \frac1\pi \left( y_0'\eta_3 + y_1'\eta_2 + y_2'\eta_1 +
y_0''(\delta_1\eta_2 +\delta_2\eta_1) + \delta_1\eta_1y_1''\right.
\nonumber\\
&&\left. {} +\frac16 y_0''' \eta_1 (3\delta_1^2 - \eta_1^2) \right),\\
\mu_3 &=& \frac1\pi \left[ y_0'(\delta_1\eta_2 + \delta_2\eta_1) +
y_1'\delta_1\eta_1 + y_0''\eta_1(\delta_1^2-\frac{\eta_1^2}{3})
+\frac{\pi^2y_0''}{6y_0'^2} \eta_1\right],\\
\nu_3 &=& \frac1{6\pi}\left( y_0'\eta_1(3\delta_1^2-\eta_1^2) - \pi^2 \frac{\eta_1}{y'_0}\right).
\eea

\end{document}